\documentclass{article}

\usepackage[english]{babel}

\usepackage[letterpaper,top=2cm,bottom=2cm,left=3cm,right=3cm,marginparwidth=1.75cm]{geometry}

\usepackage{amsmath}
\usepackage{amssymb}
\usepackage{graphicx}
\usepackage{subcaption}
\usepackage[numbers,sort&compress]{natbib}
\usepackage{tikz}
\usepackage{xspace}
\usepackage{booktabs}
\usepackage{multirow}
\usepackage{url}
\usepackage{bm}      
\usepackage{float}

\usepackage[colorlinks=true, allcolors=blue]{hyperref}

\title{Studying the Strangeness $D$-Term via Exclusive \phi Electroproduction}

\begin{document}
\newcommand{\GeVsq}{\ensuremath{\mathrm{GeV}^2}\xspace}
\newcommand{\GeV}{\ensuremath{\mathrm{GeV}}\xspace}
\newcommand{\pt}{\ensuremath{p_{T}}\xspace}
\newcommand{\Ds}{\ensuremath{D_{s}}\xspace}
\newcommand{\Dszero}{\ensuremath{D_{s}(0)}\xspace}
\newcommand{\etap}{\ensuremath{\eta'}\xspace}
\newcommand{\W}{\ensuremath{W}\xspace}
\newcommand{\MX}{\ensuremath{M_{X}}\xspace}
\newcommand{\Heep}{\ensuremath{H(e,e'p)}\xspace}
\newcommand{\abst}{\ensuremath{\lvert t \rvert}\xspace}
\newcommand{\beq}{\begin{eqnarray}}
\newcommand{\eeq}{\end{eqnarray}}
\newcommand{\la}{\langle}
\newcommand{\ra}{\rangle}
\newcommand{\PP}{\ensuremath{\mathcal{P}}}
\newcommand{\Qsq}{\ensuremath{Q^{2}}\xspace}
\newcommand{\xB}{\ensuremath{x_\mathrm{Bj}}\xspace}
\newcommand{\chisq}{\ensuremath{\chi^{2}}}
\newcommand{\hchisq}{\ensuremath{\hat\chi^{2}}}
\newcommand{\chisqA}{\ensuremath{\chi_{\rm A}^{2}}}
\newcommand{\chisqL}{\ensuremath{\chi_{\rm L}^{2}}}
\newcommand{\ndf}{\ensuremath{n_{\rm dof}}}

\newcommand{\V}{\ensuremath{\mathbb{V}}}
\newcommand{\M}{\ensuremath{\tilde{M}}}
\newcommand{\Y}{\ensuremath{Y\M}}
\newcommand{\U}{\ensuremath{\Upsilon}}

\newcommand{\dt}{\ensuremath{\bm{d}}}
\newcommand{\mb}{\ensuremath{\bm{\bar{m}}}}
\newcommand{\mt}{\ensuremath{\bm{\tilde{m}}}}
\newcommand{\s}{\ensuremath{\bm{s}}}
\newcommand{\ahut}{\ensuremath{\bm{\hat{a}}}}

\newcommand{\TODO}{\color{red} Todo.}
\newcommand{\NEW}{\color{blue}}

\definecolor{myOrange}{rgb}{1,0.5,0.}
\definecolor{myGreen}{rgb}{0.0,0.6,0.1}
\newcommand{\redtext}[1]{\textcolor{red}{#1}}
\newcommand{\bluetext}[1]{\textcolor{blue}{#1}}
\newcommand{\greentext}[1]{\textcolor{myGreen}{#1}}
\newcommand{\orangetext}[1]{\textcolor{myOrange}{#1}}

\newcommand{\commenting}[2]{\redtext{(#1: #2)}}
\newcommand{\reply}[2]{\greentext{(#1: #2)}}
\newcommand{\followup}[2]{\bluetext{(#1: #2)}}
\newcommand{\suggestion}[2]{\orangetext{(#1: #2)}}

\newcommand{\sqrts}{\ensuremath{\sqrt{s}}\xspace}
\newcommand{\emp}{\ensuremath{e^{-}p}\xspace}
\newcommand{\epp}{\ensuremath{e^{+}p}\xspace}
\newcommand{\ep}{\ensuremath{ep}\xspace}
\newcommand{\epem}{\ensuremath{e^{+}e^{-}}\xspace}

\newcommand{\invpb}{\ensuremath{\mathrm{pb}^{-1}}\xspace}
\newcommand{\alphas}{\ensuremath{\alpha_{S}}\xspace}
\newcommand{\pp}{\ensuremath{pp\xspace}}

\newcommand{\Lint}{\ensuremath{\mathcal{L}_\mathrm{int}\xspace}}

\newcommand{\pT}{\ensuremath{p_\mathrm{T}}\xspace}

\newcommand{\kT}{\ensuremath{k_\mathrm{T}}\xspace}

\newcommand{\zg}{\ensuremath{z_{\mathrm{g}}\xspace}}
\newcommand{\zcut}{\ensuremath{z_{\mathrm{cut}}\xspace}}
\newcommand{\zi}{\ensuremath{z_{i}\xspace}}
\newcommand{\zj}{\ensuremath{z_{j}\xspace}}

\newcommand{\antikT}{\ensuremath{\mathrm{anti}-k_\mathrm{T}\xspace}}

\newcommand{\Mgrsq}{\ensuremath{M_{\text{Gr.}}^{2}}}
\newcommand{\Qminsq}{\ensuremath{Q^2_\mathrm{min.}}}

\newcommand{\GIMa}{\ensuremath{M_{\text{Gr.}}^{2}/\Qminsq}\xspace}
\newcommand{\GIMln}{\ensuremath{\mathrm{ln}(M_{\text{Gr.}}^{2}/\Qminsq)}\xspace}

\newcommand{\tauoneb}{\ensuremath{\tau^{b}_{1}}}
\newcommand{\tauonebgr}{\ensuremath{\tau^{b}_{1,{\text{Gr.}}}}}

\vspace{2cm}
\begin{center}
\begin{Huge}
{\bf
Studying the Strangeness $D$-Term in Hall C via Exclusive $\phi$ Electroproduction}
\end{Huge}

\vspace{0.2cm}
\begin{Large}
{\bf
A Proposal to Jefferson Lab PAC 53
}
\end{Large}

\vspace{0cm}
\end{center}

 \vspace*{0.0cm}
    \begin{center}
        \noindent {W. Armstrong, F. A. Flor, S. Joosten\textsuperscript{*}, B. Kim, M. H. Kim, H. T. Klest\textsuperscript{*}\textsuperscript{\dag}, V. Klimenko, S. Lee, Z.-E. Meziani, C. Peng, N. Pilleux, P. E. Reimer, J. Xie, Z. Xu, M. Żurek}\\
        \vspace*{0.1cm}
        \noindent \emph{Physics Division, Argonne National Laboratory, Lemont, IL, USA} \\
        \vspace*{0.35cm}
        \noindent {A. Hoghmrtsyan, A. Mkrtchyan, H. Mkrtchyan, V. Tadevosyan} \\
        \vspace*{0.1cm}
        \noindent \emph{A. I. Alikhanyan National Science Laboratory
(Yerevan Physics Institute), Yerevan 0036, Armenia} \\
        \vspace*{0.35cm}
        \noindent {Y. Hatta} \\
        \vspace*{0.1cm}
        \noindent \emph{Brookhaven National Laboratory, Upton, NY, USA} \\
        \vspace*{0.35cm}
        \noindent {P. Markowitz, H. Szumila-Vance\textsuperscript{*}} \\
        \vspace*{0.1cm}
        \noindent \emph{Florida International University, Miami, FL, USA} \\
        \vspace*{0.35cm}
        \noindent {G. Niculescu, I. Niculescu} \\
        \vspace*{0.1cm}
        \noindent \emph{James Madison University, Harrisonburg, VA, USA} \\
        \vspace*{0.35cm}    
        \noindent {A. Camsonne, J.-P. Chen, S. Covrig Dusa, K. Dehmelt, D. Gaskell, J.-O. Hansen,\\ D. W. Higinbotham, D. Mack, M. McCaughan, A. Tadepelli} \\
        \vspace*{0.1cm}
        \noindent \emph{Jefferson Lab, Newport News, VA, USA} \\
        \vspace*{0.35cm}
        \noindent {C. Ayerbe Gayoso, C. E. Hyde, C. Ploen} \\
        \vspace*{0.1cm}
        \noindent \emph{Old Dominion University, Norfolk, VA, USA} \\
        \vspace*{0.35cm}
        \noindent {H. Atac, N. Ifat, S. Shrestha, N. Sparveris} \\
        \vspace*{0.1cm}
        \noindent \emph{Temple University, Philadelphia, PA, USA} \\
        \vspace*{0.35cm}  
        \noindent {H. Bhatt, W. Li, Z. Yin} \\
        \vspace*{0.1cm}
        \noindent \emph{Mississippi State University, Mississippi State, MS, USA} \\

        \vspace*{0.35cm}
        \noindent {M. Paolone, C. Paudel} \\
        \vspace*{0.1cm}
        \noindent \emph{New Mexico State University, Las Cruces, NM, USA} \\
     
        \vspace*{0.35cm}
        \noindent {N. Heinrich, G. Huber, M. Junaid, V. Kumar, A. Postuma, A. Usman} \\
        \vspace*{0.1cm}
        \noindent \emph{University of Regina, Regina, SK, Canada} \\
        \vspace*{0.35cm}
    
        \noindent {M. Elaasar} \\
        \vspace*{0.1cm}
        \noindent \emph{Southern University at New Orleans, New Orleans, Louisiana, USA} \\
        \vspace*{0.35cm}
                \noindent {J. Datta} \\
        \vspace*{0.1cm}
        \noindent \emph{Stony Brook University, Stony Brook, NY} \\
        \vspace*{0.35cm}
        \noindent {D. Biswas, M. Bo\"er, K. Tezgin} \\
        \vspace*{0.1cm}
        \noindent \emph{Department of Physics, Virginia Tech, Blacksburg, VA, USA} \\
        \vspace*{0.35cm}
        \noindent {S. Kay} \\
        \vspace*{0.1cm}
        \noindent \emph{University of York, York, UK} \\
           \vspace*{0.35cm}
        \noindent {D. Androić} \\
        \vspace*{0.1cm}
        \noindent \emph{University of Zagreb, Faculty of Science, Zagreb, Croatia} \\
           \vspace*{0.35cm}

        \end{center}               

\renewcommand\thefootnote{\fnsymbol{footnote}}
\footnotetext[1]{Spokesperson}
\footnotetext[2]{Contact person}

\renewcommand\thefootnote{\arabic{footnote}}
\begin{titlepage}
        \LARGE  
        \textbf{Executive Summary}
        \large 
        \vspace*{1cm}
        
        \noindent\textbf{Main Physics Goals:}\\
        \hangindent=1em
        \hangafter=1
       We propose a measurement of exclusive electroproduction of $\phi$ mesons near threshold in Hall C. We will measure the \abst-dependence of the exclusive $\phi$ electroproduction cross section, which has recently been proposed as an observable sensitive to the strangeness $D$-term~\cite{Hatta:2021can,Hatta:2025vhs}. The contribution of strangeness to the total $D$-term is presently unknown, with different arguments favoring \Ds~being large, being small, or even having opposite sign from the total $D$-term. Our exploratory measurement is designed to distinguish between these hypotheses. If \Ds turns out to be small, $\phi$ electroproduction can be used to study the gluon $D$-term. In addition, this dataset will allow us to perform measurements of other exclusive meson final states, including the first measurement of \etap electroproduction and multi-differential measurements of $\eta$ and $\omega$ electroproduction.
       \\

      \noindent\textbf{Proposed Measurement:}\\
        \hangindent=1em
        \hangafter=1
        We request 35 days of beam to measure the cross section for near-threshold deep exclusive $\phi$ production as a function of momentum transfer \abst via the missing mass of the $\Heep \phi$ reaction using the spectrometers in Hall C. We will use the SHMS at 13\textdegree~to detect electrons with a central momentum of 6.7 GeV and the HMS at 32\textdegree~to detect protons with central momentum of 1.1 GeV. The electron kinematics correspond to an average \Qsq of 3.4 GeV$^{2}$ and an average $W$ of 2.25 GeV. The observable sensitive to the strangeness $D$-term is the shape of the \abst-distribution, particularly at low-\abst, where a non-zero \Ds manifests as a softening or even an inversion of the \abst-slope. The value of \Dszero will be extracted by fitting the measured cross sections. 
        \\

        \noindent\textbf{Specific Requirements on Detectors, Targets, and Beam:}\\
        \hangindent=1em
        \hangafter=1
        This measurement will use the standard Hall C equipment, with the exception that the HMS should be connected directly to the scattering chamber with a vacuum tube to minimize multiple scattering. We propose to utilize the 10 cm liquid hydrogen target with an electron beam at an energy of 10.6 GeV and a beam current of 75 $\mu$A. \\

        \noindent\textbf{Resubmission:}\\
         \hangindent=1em
         \hangafter=1
        This proposal is an updated version of our letter-of-intent to PAC52 with the same title (LOI12-24-003). The comments of the PAC and TAC to our LOI are incorporated in this proposal. Compared to the LOI, we utilize more robust theoretical calculations that enable us to project our sensitivity to the strangeness $D$-term directly, incorporating the effect of gluons. The effects of final-state interactions and particle identification are addressed in more detail. Additionally, we now include projections for measurements of the $\etap, \eta,$ and $\omega$ electroproduction cross sections.


\end{titlepage}

\section{Introduction}
The past seven decades have seen tremendous advancement in the understanding of the electromagnetic structure of the proton, in particular via the measurement of the electromagnetic form factors. These form factors are defined through the matrix elements of the electromagnetic current operator and encapsulate the non-pointlike nature of the nucleon charge distribution. More recently the first headway has been made into understanding the \textit{mechanical} structure of the proton through the language of the QCD energy-momentum tensor (EMT). With our proposed measurement, we aim to extend this research program into uncharted territory by exploring the contribution of strangeness to the proton's mechanical structure.

The proton gravitational form factors (GFFs) encode information about the matrix elements of the energy-momentum tensor. They are often written as:

\begin{small}
\begin{equation}
\langle p'|T_{a}^{\alpha\beta}|p\rangle=\bar{u}(p')\left[A_{a}(t)\gamma^{(\alpha}P^{\beta)}+B_{a}(t)\frac{P^{(\alpha}i\sigma^{\beta)\lambda}\Delta_\lambda}{2M} + D_{a}(t)\frac{\Delta^\alpha \Delta^\beta-g^{\alpha\beta}\Delta^2}{4M} + \bar{C}_{a}(t)M g^{\alpha\beta}\right]u(p),
\label{eq:GFFs}
\end{equation}
\end{small}
where $T^{\alpha\beta}_{a}$ are the parton flavor $a$ components of the QCD energy momentum tensor ($A^{(\alpha}B^{\beta)}\equiv (A^\alpha B^\beta+A^\beta B^\alpha)/2$), $M$ is the proton mass, $P=(p+p')/2$, and $\Delta=p'-p$. The functions $A_a(t)$, $B_a(t)$, $D_a(t)$, and $\Bar{C}_a(t)$ are the gravitational form factors\footnote{Different notations for the GFFs exist in the literature, we follow the notation of Ref.~\cite{Hatta:2025vhs}. Our $B(t)$ form factor is related to the $J(t)$ form factor via $J(t)=\frac{1}{2}[A(t)+B(t)]$. $D(t)$ is occasionally denoted as $C(t)$, however they differ by a factor of four, i.e. $D(t)=4C(t)$.}. The subscript $a$ denotes the parton flavor, i.e., $D_s(t)$ represents the strangeness $D$-term. The GFFs at zero momentum transfer represent fundamental properties of the proton. These properties, including mass, spin, and the $D$-term (henceforth referred to as $D(t=0)$), describe how the proton reacts to changes in the space-time metric~\cite{Polyakov:2018zvc}. Strong constraints are placed on these properties and their corresponding gravitational form factors by symmetries and existing experimental measurements, with the $D$-term as the notable exception. The $D$-term encodes the spatial distributions of shear forces and pressure in the nucleon. The total $D$-term, $D(0)$, can be broken down into its partonic components via a sum rule,
\[ D(0) = D_g(0) + D_q(0) = D_g(0) + D_u(0) + D_d(0) + D_s(0) + \ldots.\]
At present, the only known experimental method to determine the total $D$-term of the proton is to measure all of the partonic components individually and sum them.

The total $D$-term has drawn extensive theoretical interest and assumed a central role in the Jefferson Lab physics program due to its direct connection to hadron mechanical properties. As the form factor associated with the off-diagonal components of the EMT, the $D$-term can be used to access a novel set of observables, such as the radial pressure distribution~\cite{Burkert:2018bqq}\footnote{See, however, the discussion in Refs.~\cite{Kumericki:2019ddg,Ji:2021mfb}.}, shear forces~\cite{Burkert:2021ith}, tangential and normal forces~\cite{Burkert:2021ith}, and the mechanical radius~\cite{Burkert:2023atx} of the proton. Existing experimental extractions include the light quark $D$-term $(D_{u,d})$ from DVCS data~\cite{Burkert:2018bqq} and the gluonic $D$-term $(D_g)$ from near-threshold $J/\psi$ photoproduction~\cite{Duran:2022xag}. Upcoming measurements of DVCS and $J/\psi$ production promise substantially improved precision on $D_{u,d}$ and $D_g$ in the next several years. The contribution to the pressure from parton species $a$ is defined as:
\beq
\label{Eq:pressure}
	p^a(r)=\frac{1}{6 m} \frac{1}{r^2}\frac{d}{dr} r^2\frac{d}{dr} 
	{\widetilde{D^a}(r)} 
	- m  \widetilde{\bar C^a} (r),
\eeq where $\widetilde{D^a}(r)$ and $\widetilde{\bar C^a} (r)$ are the Fourier transform of $D_{a}(t)$ and $\bar C_a(t)$, respectively. The form factor $\bar C$ is presently inaccessible to experiments. However, it is known that $\bar C_q = - \bar C_g$, meaning if all parton species are combined, the effect of $\bar C$ must cancel and the pressure distribution can be extracted without assumptions about the size of $\bar C$. One of the goals of our proposed measurement is therefore to provide the necessary missing piece required to combine quarks and gluons and more rigorously extract the proton pressure distribution, tangential and normal force distributions, and mechanical radius. On the other hand, the shear force distribution is defined as
\beq
\label{Eq:shear}
	s^a(r)= -\frac{1}{4 m}\ r \frac{d}{dr} \frac{1}{r} \frac{d}{dr}
	{\widetilde{D^a}(r)}
\eeq
and therefore can be extracted for individual parton species regardless of the value of $\bar C_a$. This means that a measurement of $D_s$ would enable an extraction of the strange shear force distribution in the proton with no assumption about the magnitude of $\Bar{C}$ or the GFFs of other parton species.

One may expect that due to the relatively small fraction of strangeness in the nucleon, \Dszero~should be negligible. However, an argument derived via QCD in the large $N_c$ limit~\cite{Goeke:2001tz,Hatta:2021can} suggests that the $D$-term should be approximately flavor-independent, (i.e. $D_u \approx D_d$) in spite of the very different contributions to the proton spin and momentum of the up and down quarks. The existing lattice data confirm this $D$-term flavor-independence prediction~\cite{Hackett:2023rif} for up and down; the up quark provides approximately 70\% of the total quark contribution to the $A$ form factor associated with longitudinal momentum and nearly 100\% to the $J$ form factor associated with spin, yet up and down quarks contribute approximately equally to the $D$-term. The lattice results of Ref.~\cite{Hackett:2023rif} indicate that $D_u = D_d = 0.56$. Extending the large $N_c$ argument to include also the third light quark species, \Dszero may still be sizable despite the small strange quark contribution to the momentum and spin structure of the proton. Additional theory calculations that suggest a non-negligible value of \Dszero were performed in Refs.~\cite{Won:2023ial,Won:2023zmf,Kim:2025pwk}. The authors applied the chiral quark-soliton model and found that $\Dszero\approx0.5D_u\approx0.5D_d$, indicating that strangeness may play a substantial role in the mechanical structure of the proton. Therefore, to fully determine the proton’s total $D$-term, the strange-quark contribution, $\Dszero$, must be measured, complementing the anticipated precision improvements in $D_{u,d}$ and $D_g$ in the coming years.

Another striking prediction from the chiral quark soliton model is that the $D$-term of sea and valence quarks should have opposite sign~\cite{Won:2023ial}. Systems with positive $D$-terms generally correspond to mechanically unstable systems~\cite{Polyakov:2018zvc}, and a result observing that strange quarks have a positive $D$-term would corroborate the naive expectation that strange quarks in the proton are an unstable system. 

Motivated by these arguments, we propose to measure $d\sigma/d\abst$ in near-threshold deep exclusive $\phi$ meson production with the goal of extracting the last remaining light quark component of the total $D$-term, thereby measuring for the first time the contribution of a non-valence quark to the mechanical properties of the proton. The goal of our experiment is to measure with enough precision to distinguish between the various hypotheses that $\Dszero \approx D_{u,d}\approx-0.5$, $\Dszero \approx 0.5D_{u,d}\approx-0.25$, $\Dszero \approx 0$, and $\Dszero > 0$. If \Dszero turns out to be small, exclusive $\phi$ electroproduction can provide a new channel through which to study $D_g$ and cross check results from $J/\psi$ production~\cite{Hatta:2025vhs}. For these reasons, adding the crucial missing piece of \Dszero will help bring the field of nucleon mechanical structure into its precision era.

\section{Theory and Kinematics}
\label{Sec:Theory}
\begin{figure}[h]
  \centering
    \includegraphics[width=0.33\linewidth]{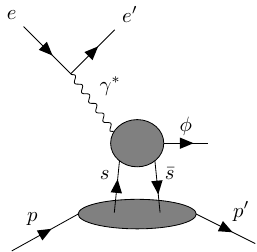}
    \includegraphics[width=0.33\linewidth]{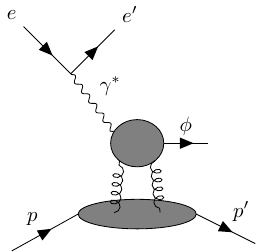}
  \caption{Leading order perturbative processes contributing to exclusive $\phi$ electroproduction. \textbf{Left:} Exclusive $\phi$ electroproduction via strange quark exchange. \textbf{Right}: Exclusive $\phi$ electroproduction via two gluon exchange.}
  \label{Fig:Feynman}
\end{figure}

Recently, the authors of Ref.~\cite{Hatta:2021can} first suggested that exclusive electroproduction of $\phi$ mesons near threshold can provide the first window into the contribution of strangeness to the total $D$-term. $\phi$ electroproduction is at present the only known observable sensitive to the strangeness $D$-term. Example diagrams contributing to this process are shown in Fig.~\ref{Fig:Feynman}. ``Near threshold" refers to the kinematic region in which $W\approx m_N + m_{\phi}$, where $W$ is the hadronic center-of-mass energy, defined as $W^2=m_N^2+2m_N(E-E') - Q^2$, with $E$ and $E'$ referring to the energies of the beam and scattered electron in the target rest frame, respectively. Near-threshold production of quarkonia can be used as a probe of the GFFs due to the relatively low momenta of the scattering hadrons in their center-of-mass frame, which enables study of long-range properties. 

A majority of the existing experimental measurements of exclusive $\phi$ production on proton targets were performed in photoproduction~\cite{Ballam:1973ph,Morrison:1979ph,Derrick:1996ph,Barth:2003phi,CLAS:2003pux,Anciant:2000kk,Klimek:2001rd,Mibe:2005phi,LEPS:2010ovn,CLAS:2013jlg,Dey:2014tfa,LEPS:2017vas,GlueX:2025ckz}, i.e., $\Qsq \approx 0$~\GeVsq. Unlike $J/\psi$ production, where the hard scale $Q^2+M_v^2$ is relatively large even for photoproduction, $\phi$ production requires $Q^2$ to be sizable to be treated within the framework of Deeply Virtual Meson Production (DVMP) factorization~\cite{Collins:1996fb}. The authors of Ref.~\cite{Hatta:2025vhs} utilize a new approach to analyzing near-threshold DVMP via the conformal partial-wave (CPaW) expansion~\cite{Mueller:2005ed}. We recount the necessary portions of the cross section calculations here.

The differential cross section for $\phi$ production by a longitudinally polarized photon can be written as 
\beq
\frac{d\sigma_L}{dt}
= \frac{2\pi^2\alpha_{em}}{(W^2-M^2)Wp_{cm}}\left((1-\xi^2)|{\cal H}|^2 -\left(\frac{t}{4M^2}+\xi^2\right)|{\cal E}|^2-2\xi^2{\rm Re}({\cal H}{\cal E}^*)\right), \label{cross}
\eeq
where ${\cal H}$ and ${\cal E}$ are the DVMP amplitudes which contain the GPDs $H^{q,g}$ and $E^{q,g}$, respectively. The skewness variable $\xi$ is defined as $\xi=\frac{p^+-p'^+}{p^++p'^+}$, where \(
p^+ \equiv \frac{1}{\sqrt{2}}(p^0 + p^3)
\)
is the light-cone “plus” component of the proton four-momentum. The DVMP amplitudes ${\cal H}$ and ${\cal E}$ have the factorized structure
\beq
\begin{pmatrix} {\cal H}(\xi,t,Q^2) \\ {\cal E}(\xi,t,Q^2)\end{pmatrix} = e_s\frac{C_Ff_\phi}{N_cQ}\sum_{a=q,g} \begin{pmatrix} H^a(x,\xi,t,\mu^2)  \\ E^a(x,\xi,t,\mu^2)\end{pmatrix} \otimes T^a\left(x,\xi,u,\frac{Q^2}{\mu^2}\right) \otimes \varphi(u,\mu^2), \label{fact}
\eeq
where $N_c=3$, $C_F=\frac{N_c^2-1}{2N_c}=\frac{4}{3}$ and $e_s=-\frac{1}{3}$ is the electric charge of the $s$-quark. $\varphi(u)$ and $f_\phi$ are the $\phi$-meson distribution amplitude and decay constant, respectively. $T^a$ are the hard scattering amplitudes and the symbol $\otimes$ represents convolutions in $-1<x<1$ and $0<u<1$. In the CPaW formalism, in moment space the amplitudes become
\beq
\begin{pmatrix} {\cal H}(\xi,t,Q^2) \\ {\cal E}(\xi,t,Q^2)\end{pmatrix} =\kappa  \sum_{j=1}^{\rm odd}\sum_{k=0}^{\rm even}\sum_a \frac{2}{\xi^{j+1}} \begin{pmatrix} H_j^a(\xi,t,\mu^2) \\ E_j^a(\xi,t,\mu^2)\end{pmatrix} T^a_{jk}(Q^2/\mu^2)\varphi_k(\mu^2), \qquad  \kappa\equiv e_s\frac{C_Ff_\phi}{N_cQ} . \label{me} 
\eeq
The threshold approximation discussed in Ref.~\cite{Hatta:2025vhs} can be summarized by the statement that
\beq
\begin{pmatrix} {\cal H}(\xi,t,Q^2) \\ {\cal E}(\xi,t,Q^2)\end{pmatrix}  \approx  \frac{2\kappa}{\xi^2} \sum_a \begin{pmatrix} H_1^a(\xi,t,\mu^2) \\ E_1^a(\xi,t,\mu^2) \end{pmatrix} T^a_{10}(Q^2/\mu^2), \label{j=1}
\eeq
where the $j=1$ amplitudes $H_1^{q,g}$ and $E_1^{q,g}$ can be written in terms of the proton GFFs via
\begin{equation}
  \begin{aligned}
    H_1^{q,g}(\xi,t,\mu^2) &= A_{q,g}(t,\mu^2) + \xi^2\,D_{q,g}(t,\mu^2)\,,\\
    E_1^{q,g}(\xi,t,\mu^2) &= B_{q,g}(t,\mu^2) - \xi^2\,D_{q,g}(t,\mu^2)\,.
  \end{aligned}
  \label{gff2}
\end{equation}

Using this approximation\footnote{Ref.~\cite{Hatta:2025vhs} performs a model study that demonstrates the accuracy of the threshold approximation.} and the NLO hard coefficients for DVMP,
\begin{equation}
\begin{split}
{\cal H}(\xi,t,Q^2) \approx \frac{2\kappa}{\xi^2} \frac{15}{2}\Biggl[\, & \left\{\alpha_s(\mu)
+\frac{\alpha_s^2(\mu)}{2\pi}\left(25.7309-2n_f+\Bigl(-\frac{131}{18}+\frac{n_f}{3}\Bigr)
\ln \frac{Q^2}{\mu^2}\right)\right\}\\[1mm]
&\quad \times \bigl(A_s(t,\mu)+\xi^2D_s(t,\mu)\bigr) \\[1mm]
&\quad +\frac{\alpha_s^2}{2\pi}\Bigl(-2.3889+\frac{2}{3}\ln \frac{Q^2}{\mu^2}\Bigr)
\sum_q \bigl(A_{q}+\xi^2 D_{q}\bigr) \\[1mm]
&\quad +\frac{3}{8}\Biggl\{\alpha_s+\frac{\alpha_s^2}{2\pi}\Bigl(13.8682-\frac{83}{18}\ln 
\frac{Q^2}{\mu^2}\Bigr)\Biggr\}\bigl(A_g+\xi^2D_g\bigr)
\Biggr],
\end{split}
\end{equation}
\begin{equation}
\begin{split}
{\cal E}(\xi,t,Q^2) \approx \frac{2\kappa}{\xi^2} \frac{15}{2}\Biggl[\, & \left\{\alpha_s(\mu)
+\frac{\alpha_s^2(\mu)}{2\pi}\Bigl(25.7309-2n_f+\Bigl(-\frac{131}{18}+\frac{n_f}{3}\Bigr)
\ln \frac{Q^2}{\mu^2}\Bigr)\right\}\\[1mm]
&\quad \times (B_s(t,\mu)-\xi^2 D_s(t,\mu))\\[1mm]
&\quad +\frac{\alpha_s^2}{2\pi}\Bigl(-2.3889+\frac{2}{3}\ln \frac{Q^2}{\mu^2}\Bigr)
\sum_q\Bigl(B_q-\xi^2 D_{q}\Bigr)\\[1mm]
&\quad +\frac{3}{8}\Biggl\{\alpha_s+\frac{\alpha_s^2}{2\pi}\Bigl(13.8682-\frac{83}{18}\ln \frac{Q^2}{\mu^2}\Bigr)\Biggr\}
\Bigl(B_g-\xi^2D_g\Bigr)
\Biggr],
\end{split}
\end{equation} 
where $n_f=4$ and $u,d,s,c$ are the active flavors. 

Since $A$ and $D$ for $u+d$ quarks and gluons are both sizable, it becomes apparent that for our chosen kinematics of $\xi\approx0.4$ there is a significant cancellation in ${\cal H} (\xi,t,Q^2)$ in the case of up quarks, down quarks, and gluons due to the terms $\bigl(A_a(t,\mu)+\xi^2D_a(t,\mu)\bigr)$ multiplying the coefficients. Since $A_{g}\approx0.4$~\cite{Hou:2019efy} and $D_{g}\approx-2$~\cite{Hackett:2023rif,Duran:2022xag}, the term $\bigl(A_{g}(t,\mu)+\xi^2D_{g}(t,\mu)\bigr)\approx0.08$ for our kinematics of $\xi\approx0.4$ is relatively small, meaning gluons contribute less to the amplitude in the near-threshold region of large $\xi$ than would be naively expected. On the other hand, since $A_s$ is known to be small (on the order of 0.03)~\cite{Hackett:2023rif,Hou:2019efy}, $D_s$ is less suppressed in its contribution to the amplitude and therefore the cross section. This feature provides the sensitivity to $D_s$ compared to $D_g$. Qualitatively, the sensitivity that the cross section exhibits to $D_s$ is around four times that of $D_g$ in our kinematics.



\begin{figure}[h]
  \centering
    \includegraphics[width=0.8\linewidth]{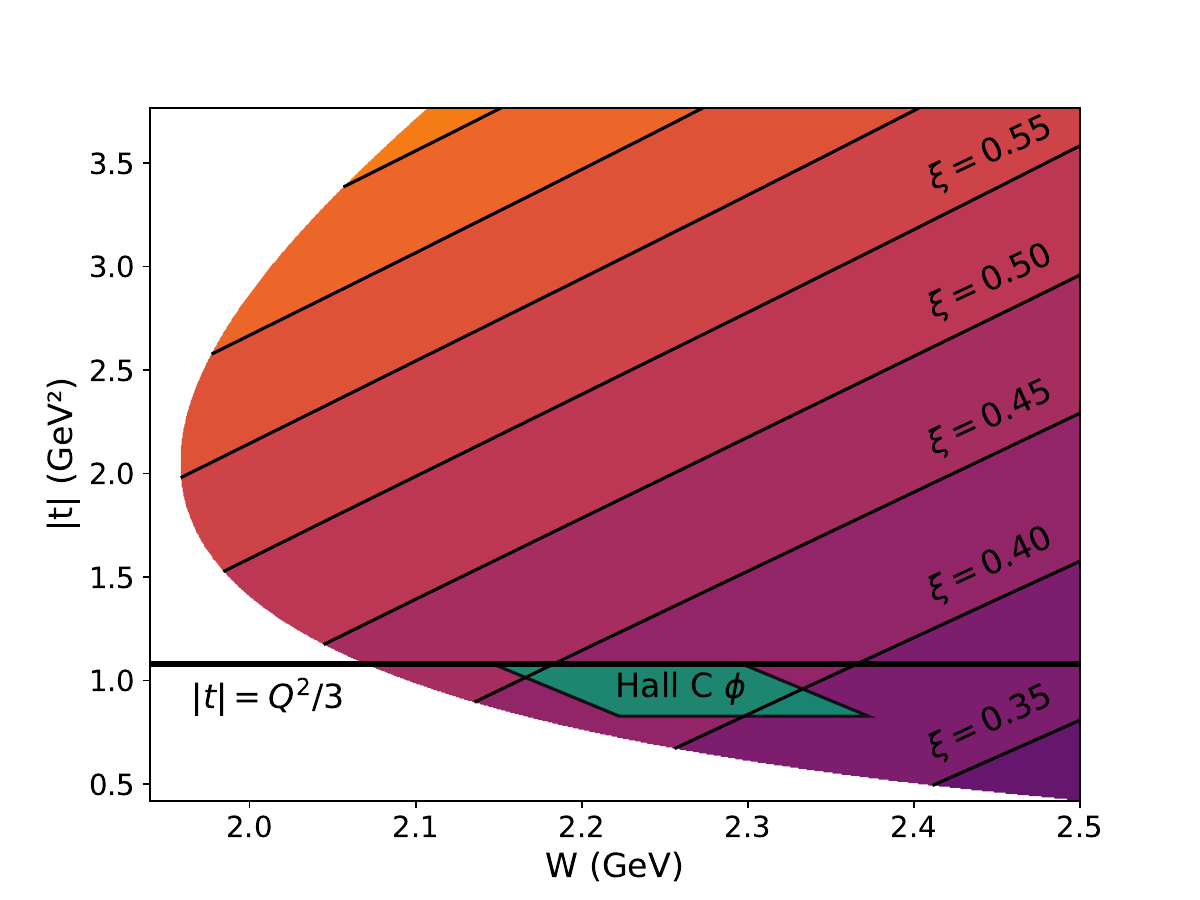}
  \caption{Contour of $|t|$ vs. $W$ with isolines of constant $\xi$ for $Q^2=3.4$ GeV$^2$. The projected Hall C $\phi$ data sit in the green trapezoidal region. To minimize the theoretical uncertainty, the data should be in the region $\xi \gtrsim 0.35$ and $|t|<Q^2/3$.}
  \label{Fig:Xi}
\end{figure}

The theory calculation outlined above provides a stringent set of kinematic requirements that should be satisfied in order to extract \Dszero. These requirements specify that the value of $Q^2$ should be a factor of 3 or more greater than $|t|$, $\xi$ should be greater than approximately 0.35, and $W$ should be close to $W_{\mathrm{Th.}}=1.96$ GeV. The kinematic region of our proposed measurement, shown in Fig.~\ref{Fig:Xi}, has been chosen to satisfy these requirements.

\begin{figure}[H]
  \centering
 \includegraphics[width=0.5\linewidth]{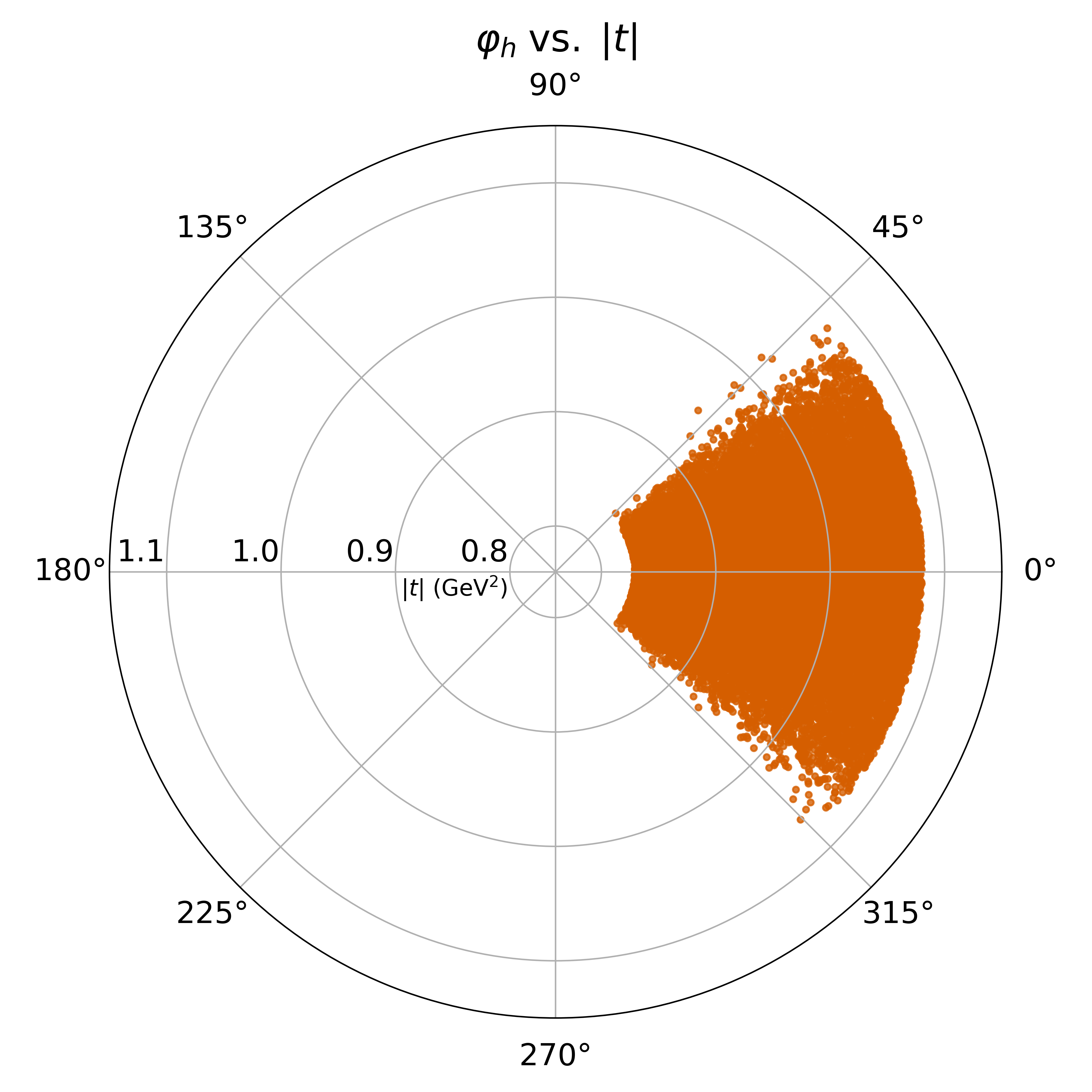} 
  \caption{Polar plot of $\varphi_h$ as a function of \abst. Larger radii correspond to larger values of \abst. The angular coverage of around 90\textdegree~is similar for all the values of \abst accessible in our measurement.}
  \label{Fig:polar_phih}
\end{figure}

Furthermore, it is important to point out that our experiment proposes to measure an unseparated $\gamma^*p$ cross section ($\sigma_0$) in a fixed window of $\varphi_h$, where $\varphi_h$ is the angle between the hadronic and leptonic planes. Our acceptance in $\varphi_h$ is shown in Fig.~\ref{Fig:polar_phih}. It should be noted that some theoretical frameworks can calculate only $\sigma_L$ while others can calculate $\sigma_0$. For comparison with the GPD predictions described above that rely on DVMP factorization, either the measured cross section must be converted into $\sigma_L$ or the theory should be augmented in some way to include $\varphi_h$ dependence. In general, the unpolarized cross section is given by 
\begin{align}
\frac{d^2\sigma_0}{dt\,d\phi_h}
&=
\underbrace{\frac{d\sigma_T}{dt}}_{\text{Transverse}}
\;+\;
\underbrace{\epsilon\,\frac{d\sigma_L}{dt}}_{\text{Longitudinal}}
\;+\;
\underbrace{\sqrt{2\,\epsilon\,(1+\epsilon)}\;\frac{d\sigma_{LT}}{dt}\,\cos\phi_h}_{\text{LT Interference}}
\;+\;
\underbrace{\epsilon\,\frac{d\sigma_{TT}}{dt}\,\cos2\phi_h}_{\text{TT Interference}}.
\end{align}

\begin{figure}[h]
  \centering
  \begin{minipage}[t]{0.48\textwidth}
    \centering
    \includegraphics[width=\linewidth]{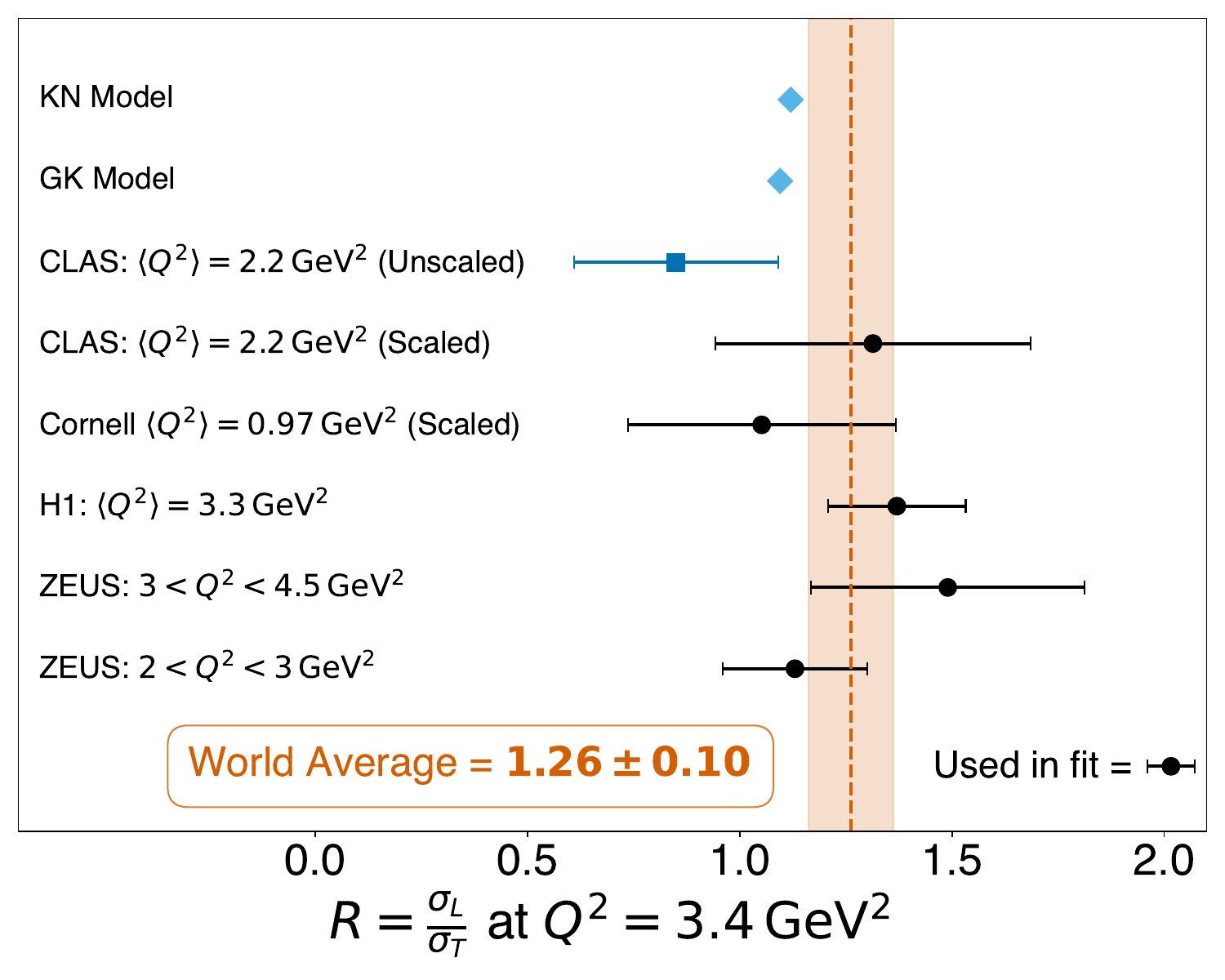}
  \end{minipage}\hfill
  \begin{minipage}[t]{0.48\textwidth}
    \centering
    \includegraphics[width=\linewidth]{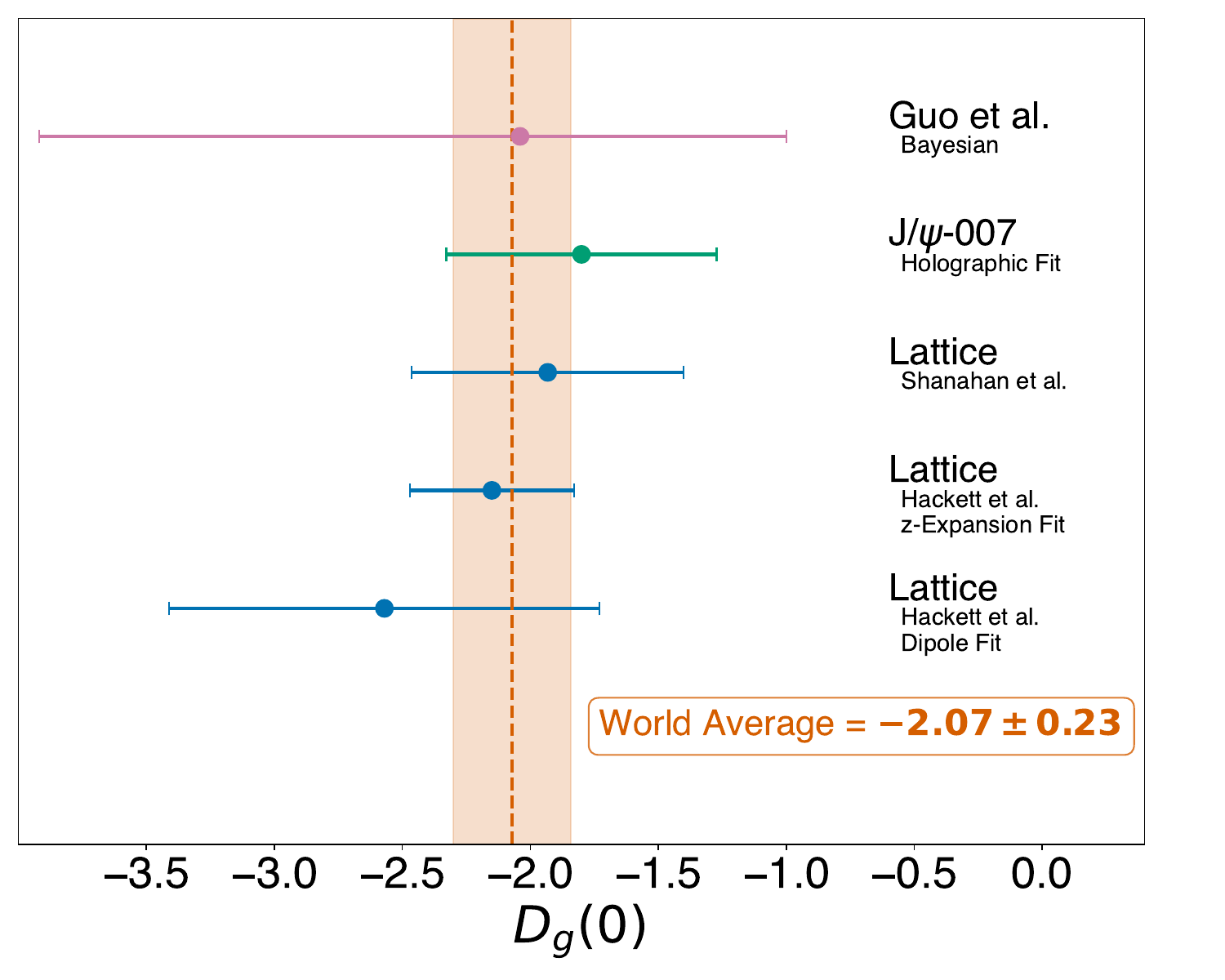}
    \label{fig:Dg}
  \end{minipage}
 \caption{\textbf{Left:} World average of $R$ at \Qsq = 3.4 \GeVsq. The CLAS and Cornell datapoints have been scaled according to the model of Ref.~\cite{CLAS12Phi}, which finds a global scaling relation between $R$ and $Q^2$ of $R(\Qsq) = 0.4\Qsq/m^2_{\phi}$. The KN model values are from Ref.~\cite{Kim:2020wrd}, and the GK model values were provided by the authors of Ref.~\cite{Goloskokov:2006hr}. \textbf{Right:} A selected weighted average of $D_g(0)$. The results are from Refs.~\cite{Guo:2025jiz,Duran:2022xag,Shanahan:2018nnv,Hackett:2023rif}\protect\footnotemark.}
 \label{fig:R}
\end{figure}

The $\varphi_h$-dependent interference terms, $\sigma_{TT}$ and $\sigma_{LT}$, should vanish if s-channel helicity conservation (SCHC) holds. Previous measurements of $\phi$ DVMP from CLAS, HERMES, H1, and ZEUS have shown that SCHC is a reasonable approximation for $\phi$ electroproduction~\cite{CLAS:2008cms,Golembiovskaya:2014cja,Augustyniak:2008gd,ZEUS:1996esk,H1:2009cml}, and its validity has been assumed by a variety of experiments. Only H1 thus far has observed a non-zero violation of SCHC in $\phi$ electroproduction~\cite{H1:2009cml,H1:2000hps}. The H1 results at $\langle Q^2\rangle =3.3~\GeVsq$ suggest that the cross section in the region of $\varphi_h$ and $\epsilon$ of our measurement is higher than the $\varphi_h$ integrated cross section by a factor of $1.20\pm0.23$. If necessary, $\sigma_{TT}$ and $\sigma_{LT}$ for $\phi$ production can be estimated for our kinematics from data\footnote{Once available, the data on $\phi$ electroproduction from Run Group A of CLAS12 in particular will be useful for estimating these terms~\cite{CLAS12Phi}.} and/or theory calculations, for example the calculations of Ref.~\cite{Kim:2020wrd}. With this knowledge of the $\varphi_h$-dependent terms, we can obtain the value of $R=\sigma_L/\sigma_T$ from existing world data and use it to convert our measured data points to $\sigma_L$ via $\sigma_L(R)=\frac{R}{1+\epsilon R}\sigma_0$\footnote{Technically, to use $R$ to determine a normalization factor, we also assume that $R$ is not a strong function of \abst in the limited range of \abst in which we measure. For the purposes of this proposal we make this assumption, which is supported by the H1 data~\cite{H1:2009cml} where no \abst-dependence of $R$ for $\phi$ production was observed. However, a \abst dependent value of $R$ can also be used to correct the data if necessary.}. In Fig.~\ref{fig:R} we perform a naive averaging of world data to provide an estimate of $R$ and its uncertainty for our kinematics. No dependence of $R$ on $W$ has been observed, and the values of $R$ from CLAS at $W\approx2.5$ GeV agree well with the data from H1 and ZEUS, which lie in the range $35 < W < 180$ GeV. In light of this observed agreement, we include the HERA collider data in our average. The resulting uncertainty on $R$ from a weighted average of existing data is 7.9\%. Propagating this uncertainty through $\sigma_L(R)=\frac{R}{1+\epsilon R}\sigma_0$, where $\langle\epsilon\rangle=0.87$ for our kinematics, produces a 3.8\% uncertainty on $\sigma_L$. We include this 3.8\% uncertainty in our study of how well \Dszero can be extracted from the projected data in Sec.~\ref{Sec:Results}. 

\footnotetext{%
    Some of the results used in our average for $D_g(0)$ are likely statistically correlated, so a weighted average is perhaps not entirely appropriate. However, we anticipate that in the coming years extractions using additional data from CLAS12 and GlueX will reach a similar precision to the O(10\%) we obtain from our averaging.
  }
The two gluon exchange process, sketched in the right panel of Fig.~\ref{Fig:Feynman}, has been suggested by various studies~\cite{Cano:2001sb,Goloskokov:2006hr,Goloskokov:2005sd,Goloskokov:2007nt} to dominate in exclusive $\phi$ production by longitudinally polarized virtual photons away from threshold, particularly at low-$x$, thereby providing access to the gluon GPD. Unavoidably, both the gluon and strange quark processes contribute to $\phi$ production, and the model of Ref.~\cite{Hatta:2025vhs} provides the necessary tools for us to disentangle these contributions. To perform that separation, the value of $D_g(0)$ needs to be included in the prediction. While we argued previously that the sensitivity of the near-threshold $\phi$ electroproduction cross section to $D_g(0)$ is less than to \Dszero, since the magnitude of $D_g(0)$ is large, it nevertheless plays a non-negligible role in the cross section. To obtain a value of $D_g(0)$ to use as an input, in Fig.~\ref{fig:R} we perform a weighted average of some existing results. The sensitivity to $D_g$ is treated as an uncertainty for the purposes of extracting \Dszero; however, in the context of a global fit this sensitivity can also provide constraints on $D_g$ from a process other than $J/\psi$ production.  

\begin{figure}[h!]
  \centering
    \includegraphics[width=\linewidth]{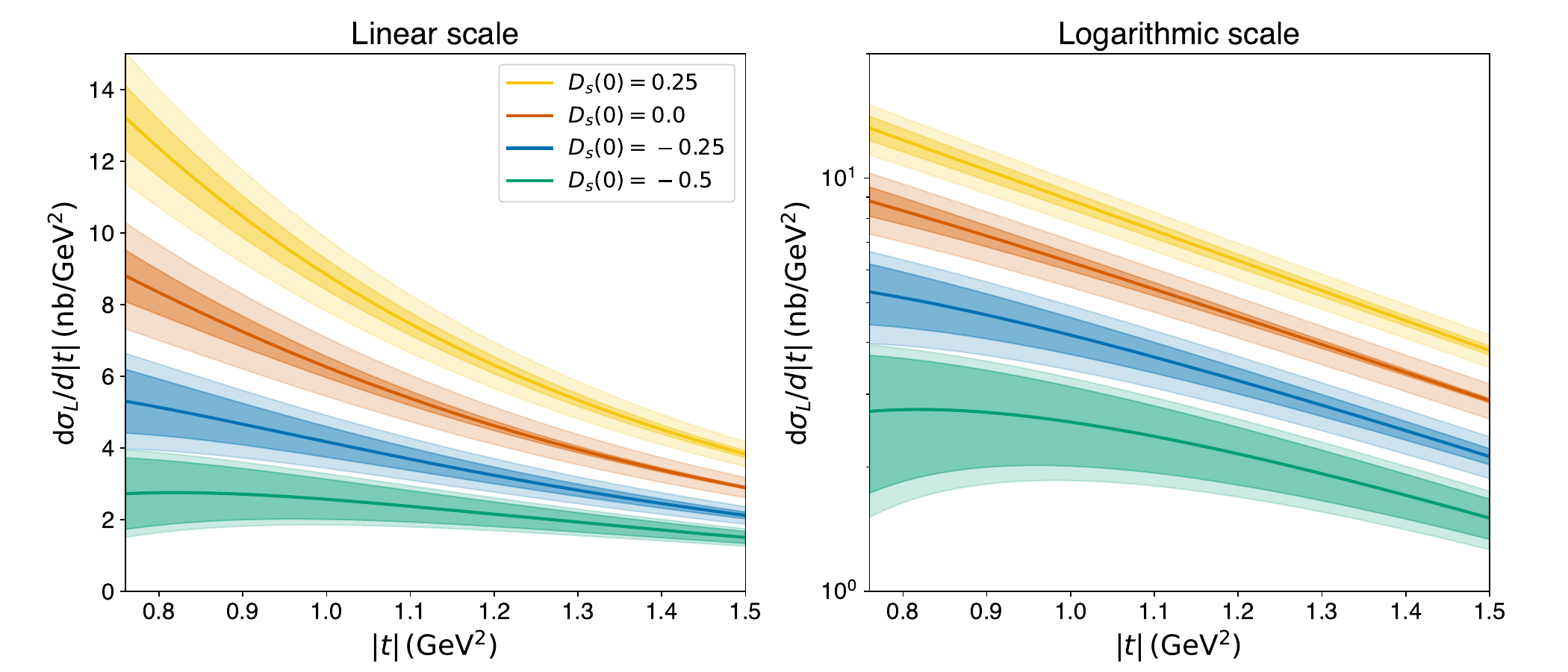}
  \caption{Theoretical predictions for $d\sigma_L/d\abst$ at \Qsq = 3.4 \GeVsq and $W=2.25$ GeV with different assumptions for \Dszero\protect\footnotemark. In this kinematic range $t_{\mathrm{min}}\approx0.8$~\GeVsq. The inner error band corresponds to the perturbative scale variation of $Q/2 < \mu < 2Q$, and the outer error band corresponds to the effect of varying $D_g(0)=-2.07$ by $\pm~0.23$ and summing it in quadrature with the scale uncertainty. Therefore, the outer error band represents the total uncertainty on the theory.}
  \label{Fig:ModelNoData}
\end{figure}
\begin{figure}[H]
  \centering
 \includegraphics[width=0.8\linewidth]{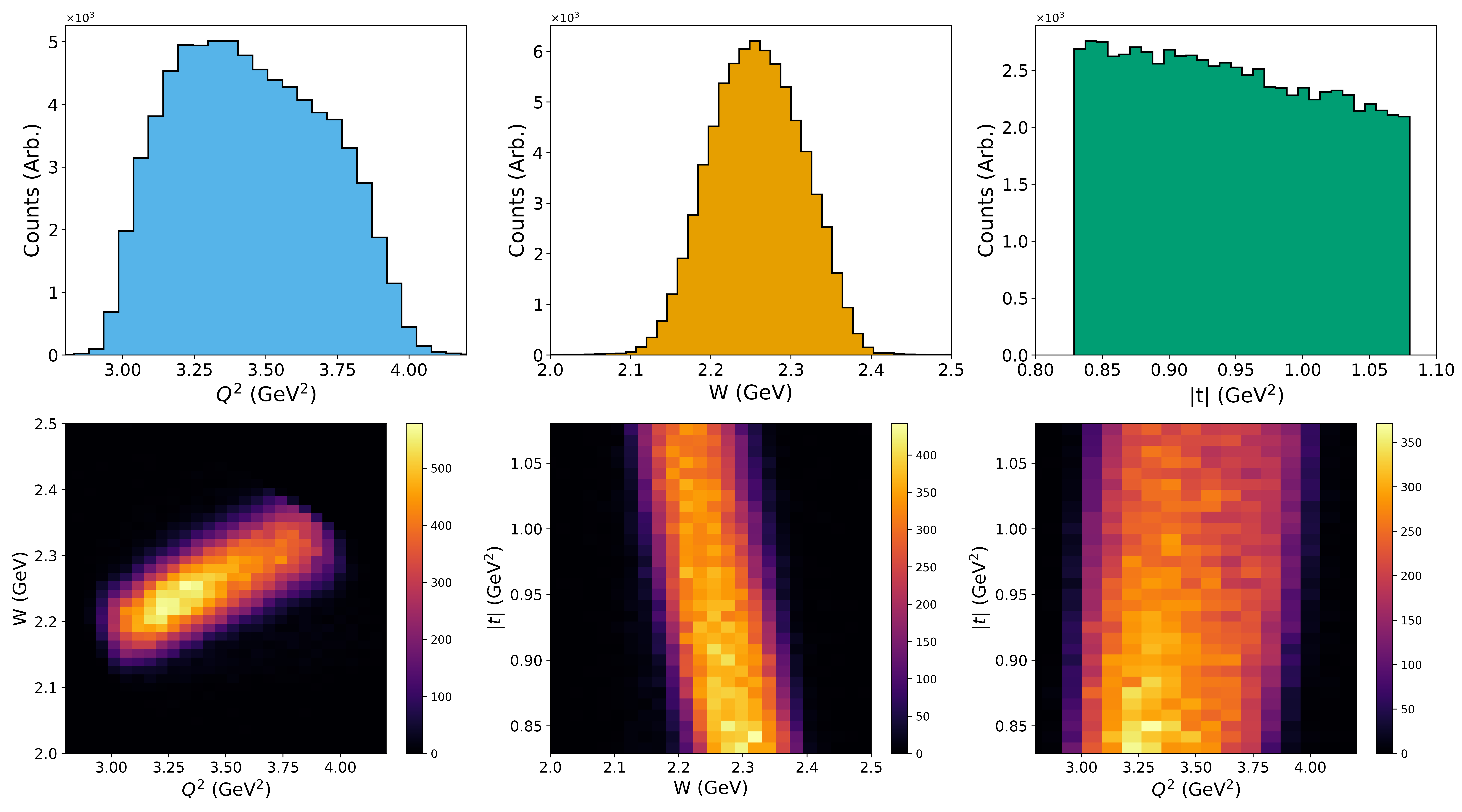} 
  \caption{\textbf{Top Row:} Reconstructed kinematics for $\phi$ events using the spectrometer settings shown in Table~\ref{Tab:Settings}. \textbf{Bottom Row:} Correlations between kinematic variables. }
  \label{Fig:kinematics}
\end{figure}

\footnotetext{
For simplicity we presently generate the theory predictions at fixed values of \Qsq and $W$, but the theory can also provide predictions integrated over the $Q^2$ and $W$ region of our experiment shown in Fig.~\ref{Fig:kinematics}.
  }

The separated $u,d,$ and $s$ quark contributions to the $D$-term were determined on the lattice in Ref.~\cite{Hackett:2023rif}. In this work, the value of \Dszero~is found to be small, with $\Ds(0) = -0.18 \pm 0.17$ and $\Ds(0) = -0.08 \pm 0.17$ for dipole and z-expansion fits, respectively. These values are consistent with zero, although the relative uncertainties are large and the extraction has not yet been performed in the continuum limit. It is also worth mentioning that the current lattice results do not conclusively preclude the possibility that \Dszero~ is positive. In intriguing scenario of $\Dszero>0$, the strange quarks would ``feel" forces in the opposite direction from the valence quarks and gluons, suggesting a fundamental difference between the mechanical properties of valence and sea quarks as suggested by Ref.~\cite{Won:2023ial}.

Given that exclusive $\phi$ electroproduction is at present the only proposed experimental observable sensitive to the strangeness $D$-term, it is evident that further theoretical and experimental input on this topic is crucial. The theory of GFFs has made exceptionally rapid progress in the last several years, motivated by the first set of experimental results. We therefore expect that by the time our experimental data analysis is complete, we will have additional models and higher precision calculations to use to extract \Dszero. An example is the holographic model of Ref.~\cite{Mamo:2021tzd}, which can calculate directly $d\sigma_L/d\abst$ and $d\sigma_T/d\abst$, thereby reducing the experimental uncertainty associated with extracting $\sigma_L$ from $\sigma_0$. For now, we provide our experimental projections for $d\sigma_L/d\abst$ and estimate our sensitivity to $\Dszero$ following the assumptions of Ref.~\cite{Hatta:2025vhs}. 

Although our emphasis throughout this proposal is on the partonic description of $\phi$ electroproduction, the kinematics of this measurement reside in a regime where both hadronic and partonic pictures have been argued to be valid~\cite{Laget:2019tou}. Near‐threshold photoproduction traditionally probes hadronic dynamics and spectroscopy, while electroproduction at moderate to high $Q^2$ naturally admits a quark–gluon interpretation. Hadronic models have, within current experimental uncertainties, described existing $\phi$-production data reasonably well~\cite{Laget:2019tou,Kim:2020wrd,CLAS:2001zwd,CLAS:2008cms}. By exploring a kinematic window in which these two descriptions overlap, our experiment can provide a test of the duality between hadronic exchanges and partonic mechanisms.

\section{Experimental Technique}
\label{Sec:Exp}
We plan to perform a measurement of $\phi$ electroproduction near threshold with a 10.6 GeV electron beam using the HMS and SHMS. The only existing measurements of $\phi$ electroproduction in the region of $W<3$ GeV were performed by CLAS~\cite{CLAS:2008cms,CLAS:2001zwd} at beam energies of 4.2 and 5.8 GeV. Unfortunately, these data cannot presently be used to study \Dszero since they do not satisfy the kinematic constraints of the theory described in Sec.~\ref{Sec:Theory} and the cross sections are measured only single-differentially with large bins. Since the shape of the \abst-distribution near threshold is sensitive to $W$ and \Qsq, an extraction of \Dszero~via the \abst-distribution should seek to measure $d\sigma/d\abst$ in a small region of \Qsq and $W$ if possible. The smallness of the cross section for $\phi$ production in the near-threshold region of interest for the study of GFFs necessitates high luminosity. The Hall C spectrometers, with their narrow angular and momentum acceptance but high luminosity capability, are therefore a natural choice for such a measurement.


The high precision of the Hall C spectrometers allows for measurements of cross sections via the missing mass technique~\cite{JeffersonLabFp:2019gpp}. We propose to leverage this capability to reconstruct the $\phi$ meson in the missing mass distribution of the $\Heep X$ reaction. The missing mass (\MX) is reconstructed as the mass of the four-vector defined by $(\Vec{P}_e + \Vec{P}_p) - (\Vec{P}_{e'} + \Vec{P}_{p'})$. This technique has some advantages over reconstructing the full final state. The first is that all decay modes of the $\phi$ contribute, i.e., there is no reduction in the measured event yield due to the branching fraction. Another advantage is the fact that identifying the scattered electron and proton is substantially easier than identifying kaons from the $\phi$ decay, since kaons are easily confused for the more copiously produced pions and protons. The primary disadvantage of the missing mass technique is that other processes inevitably will produce similar $e'p'$ missing masses to the exclusive $\phi$ events. This results in a large and partially irreducible background that must be subtracted to recover the true $\phi$ cross section. We describe the background in detail in Section~\ref{sec:bkgds}.

To reach low values of \abst where the cross section exhibits the most sensitivity to \Dszero, the HMS will be used to detect the proton and the electron will be detected in the SHMS. Since our goal is to fit the \abst-distribution in a small region of $Q^2$ and $W$ to study \Dszero, we propose only a single setting for the spectrometers. Our setting is presented in Table~\ref{Tab:Settings}. The HMS has been used successfully in similar settings with relatively low proton momenta by the VCS experiment (E12-15-001). The loss of protons to inelastic collisions with materials prior to and inside of the HMS detector hut was estimated for that experiment to be around $5\pm0.5\%$ for protons of momenta 0.8 to 0.9 GeV~\cite{Li2022}. Furthermore, the $N$-to-$\Delta$ proposal to PAC 50~\cite{NDelta} demonstrated the feasibility of detecting protons with momenta of around 0.4 GeV in the HMS, so we foresee no technical issues with our proton momentum setting of 1.1 GeV. 

\begin{table}[h!]
\centering
\begin{tabular}{cccccc}
\toprule
$P_{e'}$ \text{SHMS} & $\theta_{e'}$ \text{SHMS} & $P_{p'}$ \text{HMS} & $\theta_{p'}$ \text{HMS} & \text{Beam Current} $(\mu A)$ \\
\midrule
 6.7 GeV & 13\textdegree & 1.1 GeV & 32\textdegree & 75\\
\bottomrule
\end{tabular}
\caption{Proposed spectrometer setting and beam current, assuming the standard 10 cm LH$_2$ target. The SHMS setting corresponds to $\Qsq\approx3.4$ \GeVsq and $W\approx2.25$ GeV for exclusive $\phi$ events. The accepted kinematics for $\phi$ events is shown in Fig.~\ref{Fig:kinematics}.}\label{Tab:Settings}
\end{table}
Since the rate of negatively charged pions is a factor of 100 smaller than electrons in the SHMS at our setting, we can operate the SHMS without the Noble Gas Cherenkov to reduce multiple scattering before the SHMS drift chambers and thereby improve the missing mass resolution. Ideally the region nominally occupied by the NGC would be replaced by a snout to extend the spectrometer vacuum, but if this is not feasible a helium bag would suffice. We also intend to connect the HMS directly to the scattering chamber via a vacuum tube to improve the missing mass resolution. This modification eliminates the detrimental effects of multiple scattering in the windows of the scattering chamber and spectrometer, as well as the air in-between. Since we do not plan to move the HMS during our experiment, this tube could remain in place for the entire allotted beam time. The SHMS can remain uncoupled without significantly degrading the $M_x$ resolution\footnote{Having the flexibility to move the SHMS would allow us to compensate for unexpected changes in the beam energy during the run without significant loss of uptime.}. We simulate this configuration in our projections by removing the scattering chamber window, air, and entrance window between the target and the HMS. To ensure the spectrometer resolutions assumed in our proposal can be attained, we allocate three days for optics and calibration runs. Elastic \Heep data can be taken for calibration at 4.4 GeV with the HMS parked in place at 32 degrees, meaning the work of coupling the HMS to the scattering chamber could occur prior to the beginning of the experiment and would not need to interrupt beam delivery.


Our proposed measurement is not particularly sensitive to the central value of the beam energy. A 5-pass beam energy lower than the 10.6 GeV we assume for our projections would be acceptable, although higher would be preferred if possible. However, to avoid degrading the missing mass resolution on the $\phi$ peak, the beam $dp/p$ at the target must be smaller than $8\cdot10^{-4}$ on average. 

To simulate the exclusive $\phi$ channel, we interfaced the cross section parameterization developed for the CLAS12 exclusive $\phi$ proposal~\cite{CLAS12Phi} and the theoretical predictions of Ref.~\cite{Hatta:2025vhs} to the lAger event generator~\cite{lager_eicweb}. The CLAS12 parameterization successfully reproduces the existing world data on $\sigma_T, \sigma_L$, and $R$~\cite{CLAS:2008cms,CLAS:2001zwd,ZEUS:2005bhf,H1:2009cml,Dixon:1978vy,2000hermes}. We generate events according to whichever of the two models has a lower cross section at a given generated $W$ and $Q^2$. For the \abst dependence, we assume a tripole form, motivated by the QCD calculations of Refs.~\cite{Tong:2021ctu,Tanaka:2018wea,Lepage:1980fj}. The mass term of the tripole is taken to be 1.1 GeV, consistent with the results of Refs.~\cite{Duran:2022xag} and~\cite{Burkert:2018bqq} for $m_D$. A more complete description of this parameterization is provided in the appendix of this proposal [\ref{Sec:Appendix}]. 

To evaluate the acceptance and resolution of the Hall C spectrometers, we utilized the standard Hall C Monte Carlo program, SIMC~\cite{SIMCMonteCarlo}. The version of SIMC used for our simulations was tuned to accurately describe the $J/\psi-007$ data. Our setting was determined by maximizing the signal-to-background for exclusive $\phi$ in the relevant region of \abst while keeping \Qsq reasonably high and $W$ near the threshold value of 1.96 GeV. Reaching significantly higher values of $Q^2$ seems infeasible with the missing mass method due to the unfavorable $Q^2$ scaling of the signal compared to the DIS background. The spectrometer setting was optimized to reduce the random coincidence and physics event backgrounds as much as possible.

\subsection{Backgrounds}
\label{sec:bkgds}
The two primary physics backgrounds are exclusive production of non-$\phi$ mesons and continuum processes, including multi-pion production and DIS. These physics processes will unavoidably produce a background in the missing mass distribution upon which the $\phi$ peak will sit, and understanding this background is a vital component of this experiment. The masses and widths of some of the relevant mesons are reproduced in Table~\ref{Tab:Mesons}. 
\begin{table}[h!]\centering
\begin{tabular}{lcc}
\toprule
Meson & Mass (MeV) & Width (MeV) \\
\midrule
\textbf{Vector Mesons} & & \\
$\rho$ & 775.3 & 149.1 \\
$\omega$ & 782.7 & 8.68 \\
$\phi$ & 1019.46 & 4.25 \\
\midrule
\textbf{Pseudoscalar Mesons} & & \\
$\pi$ & 134.98 & small \\
$\eta$ & 547.86 & 0.00131 \\
$\eta'$ & 957.78 & 0.196 \\
\bottomrule
\end{tabular}
\caption{Masses and widths of vector and pseudoscalar mesons (in MeV) relevant for this measurement~\cite{ParticleDataGroup:2024cfk}.}
\label{Tab:Mesons}
\end{table}

\begin{figure}[h]
  \centering
    \includegraphics[width=0.43\linewidth]{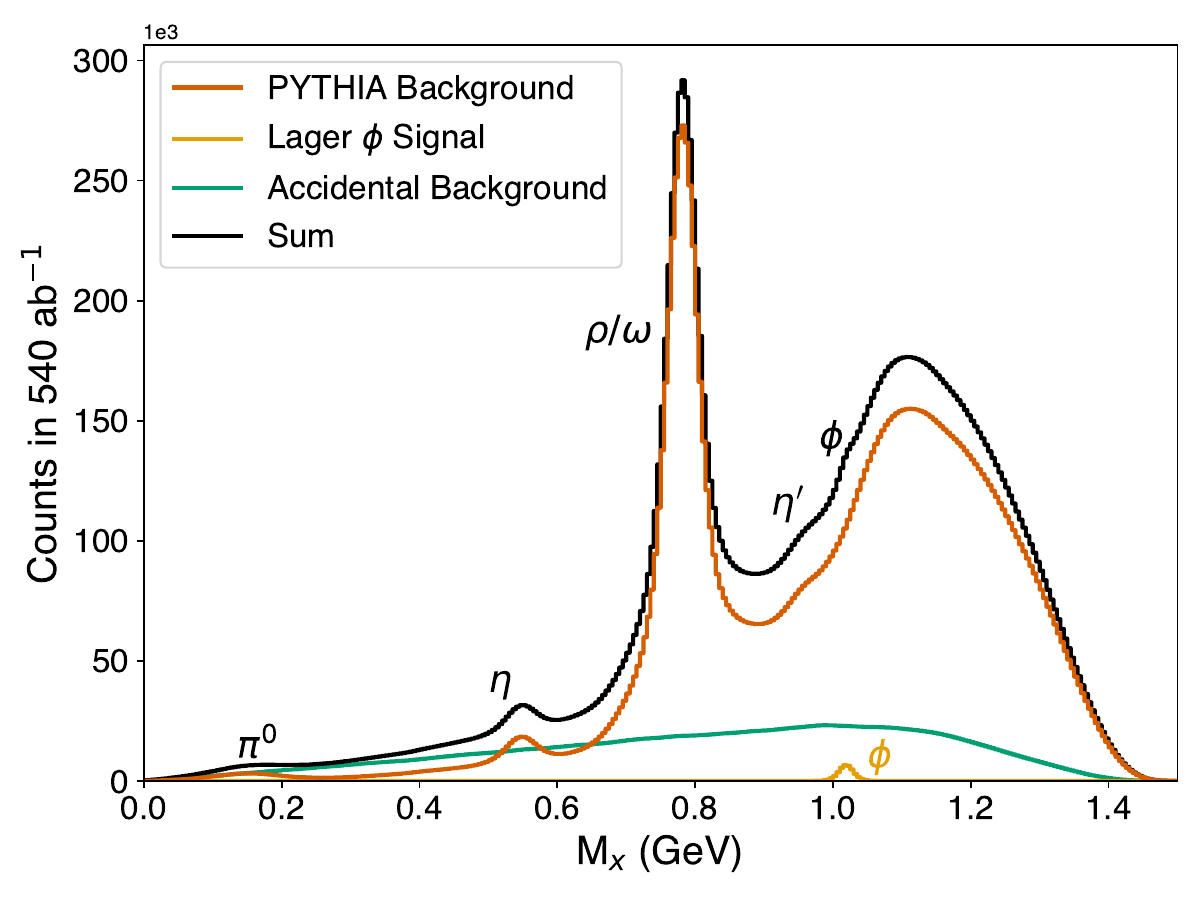}
    \includegraphics[width=0.46\linewidth]{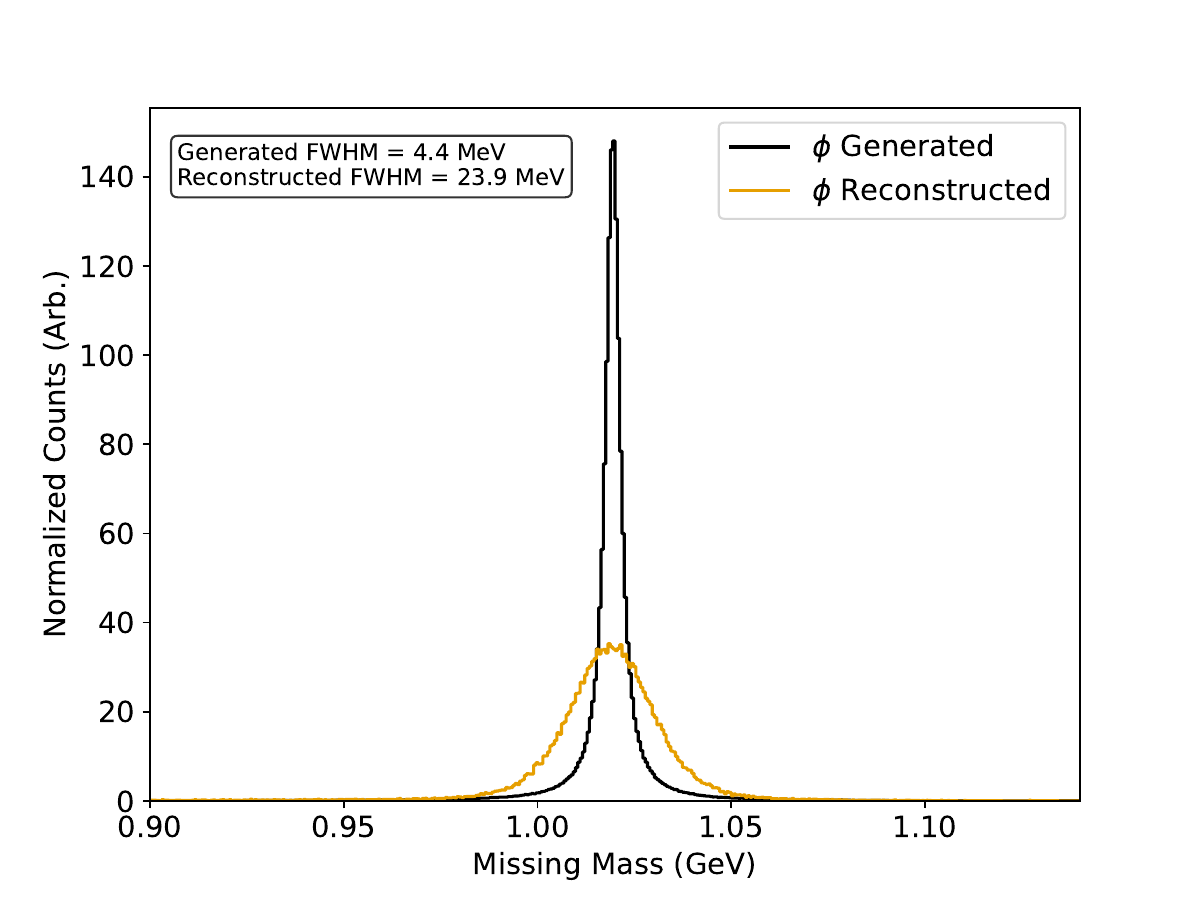}
  \caption{\textbf{Left:} Reconstructed missing mass $(\MX)$ spectra of 540 ab$^{-1}$ of events. The peaks corresponding to the $\pi^0, \eta, \rho, \omega,$ and \etap are all visible. \textbf{Right}: Reconstructed and true missing mass $\MX$ spectra for $\phi$ events.}
  \label{Fig:Mx}
\end{figure}

The general-purpose MC generators PYTHIA6 (PYTHIA eRHIC tune) and LEPTO (CLASDIS tune) were used to evaluate the contributions to the background from exclusive production of mesons other than the $\phi$, as well as DIS and other continuum processes. The cross section of the continuum background was nearly identical in both PYTHIA and LEPTO in the region underneath the $\phi$ peak. The reconstructed \MX distribution from PYTHIA6 is shown in Fig.~\ref{Fig:Mx}. The cross section for the $\eta'$ has never been measured in electroproduction, and its proximity to the $\phi$, only 62 MeV away, makes it an important background to understand. The \etap has been previously observed in the missing mass spectrum during the 6 GeV $u$-channel $\omega$ $\Heep \omega$ analysis of Refs.~\cite{JeffersonLabFp:2019gpp,Li:2017xcf}. The existing photoproduction data suggest that at $W\approx2.25$ GeV, the $\etap$ cross section is a factor of two larger than the $\phi$. Ref.~\cite{Goloskokov:2011rd} has made predictions for the \etap cross section using a GPD model, and the authors produced predictions for our kinematics. It is known that PYTHIA6 natively overestimates the \etap cross section, and therefore an \etap suppression factor is included as a tunable parameter in the program. We nominally take the value of this parameter such that the \etap cross section in PYTHIA approximately agrees with the predictions provided to us by the authors of Ref.~\cite{Goloskokov:2011rd}. 

PYTHIA eRHIC has an interface to RADGEN~\cite{Akushevich:1998ft} to enable predictions with electromagnetic radiative corrections. The background predictions shown above did not use the radiative option, but we studied the differences between the radiative and non-radiative predictions to assess the overall impact of internal QED radiation. The emission of a photon off the electron in the initial- or final-state typically increases the missing mass of an event, thereby modifying slightly the background and signal distributions. Inclusion of internal QED radiation results in a few percent higher continuum background arising from exclusive meson or other low-\MX events being shifted to higher \MX, and we expect a similar effect for $\phi$ events. The $\phi$ events lost to QED radiation will be corrected for as part of the radiative correction, to which we assign a 4\% uncertainty. We independently assessed the contribution of radiative $ep$ elastic events to the background using the ESEPP generator~\cite{esepp}. Approximately 0.2 events per ab$^{-1}$ were reconstructed. We therefore expect on the order of 100 background events from the radiative elastic $ep$ scattering process, which will have a negligible impact on the measurement.

The rate of single charged particles entering the spectrometer acceptances was determined using PYTHIA6 with no cuts on any kinematic variables and independently via the Wiser parameterization. We estimate based on the thickness and density of the aluminum target windows that around 10\% of the rate will originate from the target windows. PYTHIA and Wiser both predict that the rate of random coincidences between a negatively charged particle in the SHMS acceptance and a positively charged particle in the HMS acceptance is around 650 Hz for a trigger coincidence time window of 70 ns. Assuming the trigger is formed by a coincidence of the HMS and SHMS hodoscopes, the rates are well within the capabilities of the data acquisition system and the livetime should be close to 100\%. The rate of protons in the HMS is around 165 kHz, and the rate of electrons in the SHMS is around 25 kHz. The relevant singles rates are given in Tab.~\ref{Tab:SinglesRates}. The central momentum and angle of the HMS setting were chosen in part to minimize the singles rate. 

\begin{table}[h!]
\centering
\resizebox{\columnwidth}{!}{%
  \begin{tabular}{lccccc}
  \toprule
   & \text{Total Rate SHMS} & \text{$e^{-}$ Rate SHMS} & \text{Total Rate HMS} & \text{p$^+$ Rate HMS}  & \text{K$^+$ Rate HMS} \\
  \midrule
   PYTHIA6 & 23 kHz & 22.5 kHz & 380 kHz & 160 kHz & 14 kHz\\
  \midrule
   Wiser   & 26 kHz & 25 kHz & 370 kHz & 170 kHz & 21 kHz\\
  \bottomrule
  \end{tabular}%
}
\caption{Spectrometer singles rates, determined with PYTHIA6 and compared to the Wiser parameterization. The PYTHIA values are determined for the LH$_2$ target and scaled up by a factor of 10\% to account for the effect of the target windows.}
\label{Tab:SinglesRates}
\end{table}

We now consider the issue of particle identification, as suggested by the TAC report to our letter-of-intent to PAC52~\cite{Klest:2025rek}. The scattered electron measured in the SHMS can provide a start time to which the time-of-arrival of hadrons at the hodoscopes of the HMS can be compared. The coincidence time resolution of the SHMS and HMS is around 200 ps. The difference in arrival time at the HMS hodoscopes between protons and kaons with 1.1 GeV of momentum is more than 10 nanoseconds, meaning contamination from hadrons other than protons arising from the same beam bunch as the scattered electron can be safely ignored. 

\begin{figure}[h]
  \centering
  \includegraphics[width=0.98\linewidth]{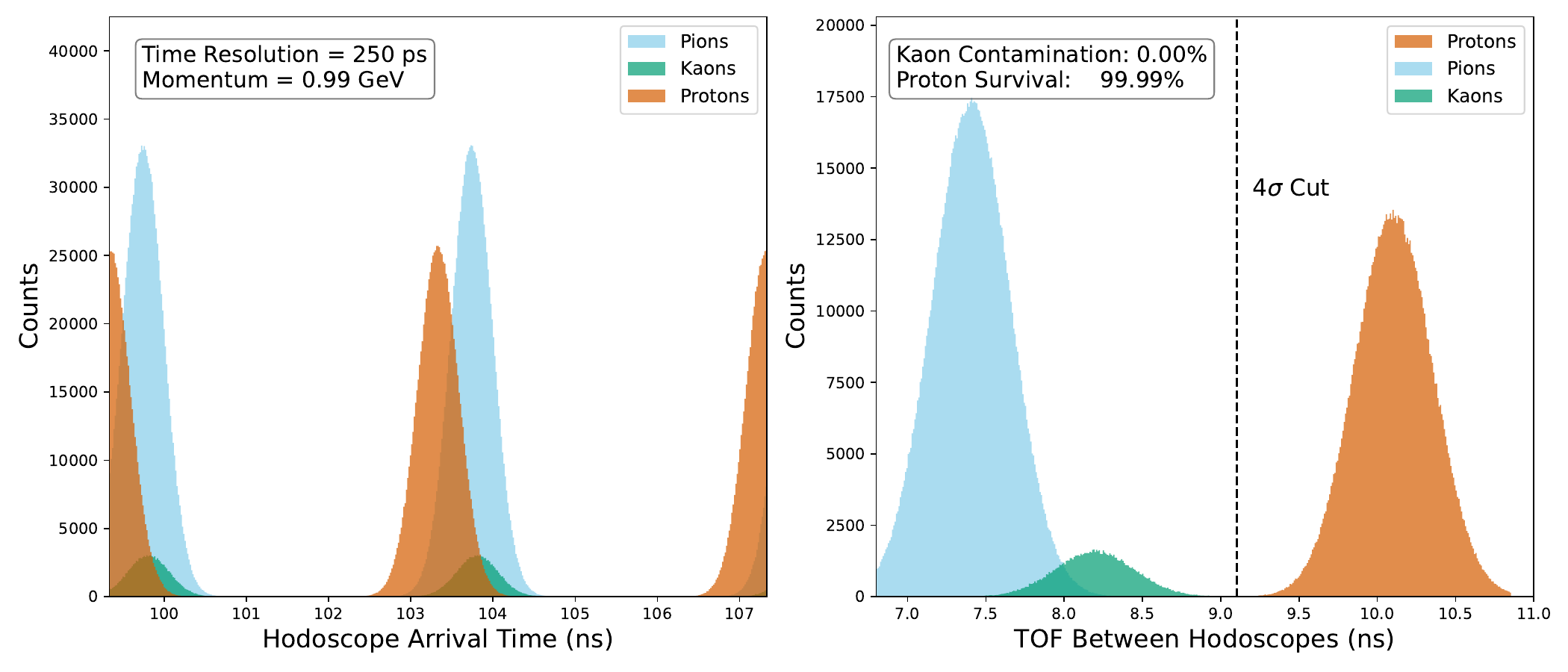}
  \vspace{0.5em}
  \includegraphics[width=0.98\linewidth]{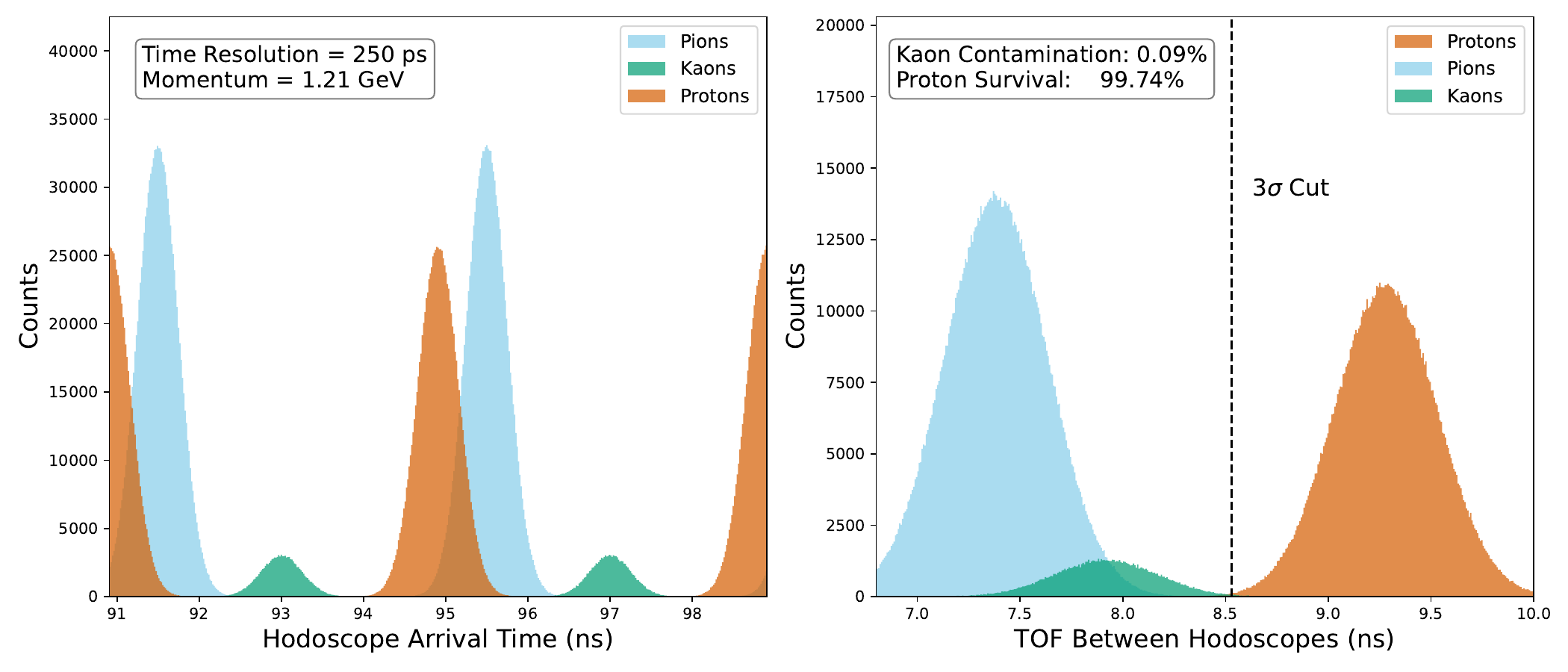}
  \caption{%
    \textbf{Top Left:} Time‐of‐arrival distributions at 250 ps resolution for hadrons with 0.99 GeV of momentum ($\delta=-10\%$), showing pion, kaon, and proton arrival times at the HMS hodoscopes. \textbf{Top Right:} $\Delta t$ between the HMS hodoscope planes. 
    \textbf{Bottom:} Same as above for 1.21 GeV ($\delta=+10\%$). While the separation from the $\Delta t$ between the hodoscope planes diminishes, the number of kaons arriving at the hodoscopes at similar times as protons also decreases.
    The relative fluxes are taken from Table~\ref{Tab:SinglesRates}.}
  \label{Fig:TOFPID250ps}
\end{figure}

While confusion of protons for other hadrons from the same beam bunch is negligible, there can still be random coincidences of an electron and a proton from the same beam bunch, or a coincidence between an electron and a pion or kaon from a different beam bunch. The differences in the time-of-flight from the target to the detector hut assuming a baseline of 22.5 m for protons, pions, and kaons are large enough such that there are wrap-around ambiguities with respect to the RF time. In fact, the scattered protons from our events of interest will arrive at the first HMS hodoscope at similar times as kaons produced four RF periods later and pions produced five RF periods later, assuming operation at 250 MHz. The left panel of Fig.~\ref{Fig:TOFPID250ps} shows approximately what the time-of-arrival spectrum will look like. For 250 MHz operation, protons and kaons with 1.21 GeV of momentum are well-separated in time-of-arrival, but at lower momenta they can overlap. The ambiguity between species can be lifted if the particle species can be identified in the HMS via the time-of-flight between the two HMS hodoscope stations. The distance between the two hodoscope planes of the HMS is 220 cm, and the resolution on the time-of-flight between them is approximately 200 picoseconds\footnote{The effect on the time-of-flight from multiple scattering in the HMS detector materials between the two hodoscope planes was studied and found to be negligible. Protons see an average delay with respect to their nominal time-of-flight of 23 ps with a $\sigma$ of 1.2 ps from multiple scattering while kaons see an average delay of 6 ps with a $\sigma$ of 0.3 ps.}. The right panels of Fig.~\ref{Fig:TOFPID250ps} show that protons can reliably be separated from kaons at all momenta accessible in our setting using the time-of-flight between the two hodoscope planes with 250 ps of timing resolution. 


Now we assess the impact of $e+p$ random coincidences from the same beam bunch on the offline analysis. After analysis cuts, the random coincidence rate of electrons and protons from the same RF bucket is $\approx$~5.5 Hz if no cuts are applied on the vertex location along the target. The target position resolutions of the spectrometers are on the order of 0.3 cm, meaning electrons and protons with sufficiently different measured vertex positions can be rejected. Assuming a cut on the difference between the vertex location as measured by the HMS and SHMS, the random coincidence rate can be reduced by a factor of 3 with minimal loss of signal. The remaining rate of 1.8 Hz is nevertheless not negligible compared to the physics background event rate of $\approx$~6 Hz and the exclusive $\phi$ event rate of $\approx$~0.02 Hz; however most random coincidence events fall outside the $M_x$ region of interest. After analysis cuts, the rate of random coincidence events in the $\phi$ peak region is around 0.07 Hz, compared to 0.32 Hz for the physics background. Therefore, the background from random coincidences is estimated to be $\approx25\%$ of the total background. This additional 25\% background, shown in green in Fig.~\ref{Fig:Mx}, is incorporated into the signal extraction procedure to account for random coincidences. 

\subsection{Final-state Interactions}
\label{sec:FSI}
The PAC report for our letter-of-intent for this measurement highlighted that final-state interactions between the produced $\phi$ and scattered proton should be understood for a full proposal. The final-state interaction between the $\phi$ and the proton was calculated for $\phi$ photoproduction in Ref.~\cite{Kim:2024lis} and determined to be approximately two orders of magnitude smaller than the already small production cross section in the kinematic region of $W\approx2.25$ GeV. We therefore neglect final-state interactions between the produced $\phi$ before its decay and the proton. 

Rescattering of the $\phi$ decay products on the proton, thereby altering the measured $M_x$, poses a larger challenge. We can estimate that effect using the $\phi$ lifetime and the cross sections for the decay products to interact with the proton. We performed a simulation using our generated sample of exclusive $\phi$ events to determine the effect on the $M_x$ distribution from kaon and pion rescattering. The $c\tau$ of the $\phi$ is 46.5 fm, and the primary decay modes are $\phi\rightarrow K^+K^-, \phi\rightarrow K^0_LK^0_S,$ and $\phi\rightarrow \pi^+\pi^-\pi^0$, with branching fractions of 48.9\%, 34.2\%, and 15.3\%, respectively~\cite{ParticleDataGroup:2024cfk}. The total elastic and inelastic cross sections for $K+p$ and $\pi+p$ interactions at these energies are reasonably well known from a large body of world data~\cite{Peresedov:2019vrx,Falk-Vairant:1960ora,Arndt:2003fj}. We assume a conservative estimate for the elastic scattering cross section in the range of $K/\pi+p$ center-of-mass energies for our kinematics of 100 mb and, for the inelastic case, 20 mb\footnote{The elastic and inelastic cross sections assumed here are around a factor of 2 larger than the measured values.}. We take the $K^0_{L,S}+p$ cross sections to be the same as $K^-+p$. Treating these cross sections as black disks with radii $r_{int.}$ defined via $\sigma=\pi r_{int.}^2$, we determine the rescattering probability as the likelihood that the proton and $\phi$ decay products come within a distance of $r_{int.}$ of one another. For cross sections of 100 mb, $r_{int.}$ is 1.8 fm and the likelihood of elastic rescattering is roughly 2\%. For the inelastic case, the rescattering probability is 0.5\%. These probabilities are primarily controlled by the $\phi$ lifetime. For our kinematics, the $\phi$ is fairly boosted and thus survives long enough before decaying that the proton and decay products typically do not interact. 

For elastic scattering, the functional form of $d\sigma/d|t|$ can be approximated as $e^{-B|t|}$. The values of $B$ for $p+K$ and $p+\pi$ interactions have been measured at a variety of energies and are around 1 GeV$^{-2}$ in our kinematics. For events which do experience elastic rescattering, the average change in $M_x$ is around 0.1 GeV, meaning most of these events will likely be lost outside the region of the $\phi$ peak and will require a correction. It is also assumed that events which experience inelastic rescattering will not be reconstructed. Since the above analysis represents a naïve estimate, we include a 2\% point-to-point uncertainty on the data points associated with the rescattering correction.

\subsection{Projected Uncertainty}

Due to the relatively small $\phi$ cross section, it is expected that the extraction of the $\phi$ signal from the background will prove to be the largest source of experimental uncertainty. For this measurement, higher integrated luminosity will not only increase the signal yield but also reduce the statistical fluctuations in the background and enable a more reliable signal extraction. To measure the shape of the \abst distribution with reasonable precision, we assume six equal-sized bins between $0.828 < \abst < 1.078~\GeVsq$ with a width of roughly 0.042 \GeVsq each. Protons with 1.1 GeV of momentum in the HMS have a \abst resolution of around 0.0028 GeV$^2$, which is well below our bin size. 

The impact of finite statistical precision on the $\phi$ signal extraction procedure was evaluated using a replica method. In each replica, the simulated $\phi$ signal and simulated background data points in the missing mass distribution are scattered in accordance with their statistical uncertainty and the signal extraction is performed. In this case, the signal extraction involves fitting a polynomial to the background in the sideband region around the $\phi$ and subtracting the resulting curve from the measured event yield (signal plus background). Due to the computational challenge of generating a full background dataset, we simulated roughly three days (54 ab$^{-1}$) of background events and used the shapes of those distributions to parameterize the background in each \abst bin. The parameterization enabled sampling the background distribution with different assumed integrated luminosities and examining the effects on the stability of the signal extraction. In the actual experiment, events which kinematically could not have originated from an exclusive $\phi$, e.g., $W<1.96$ GeV, can be used for a data-driven determination of the background underneath the $\phi$ peak.

\begin{figure}[t]
  \centering
  \begin{subfigure}[b]{0.49\textwidth}
    \includegraphics[width=\linewidth]{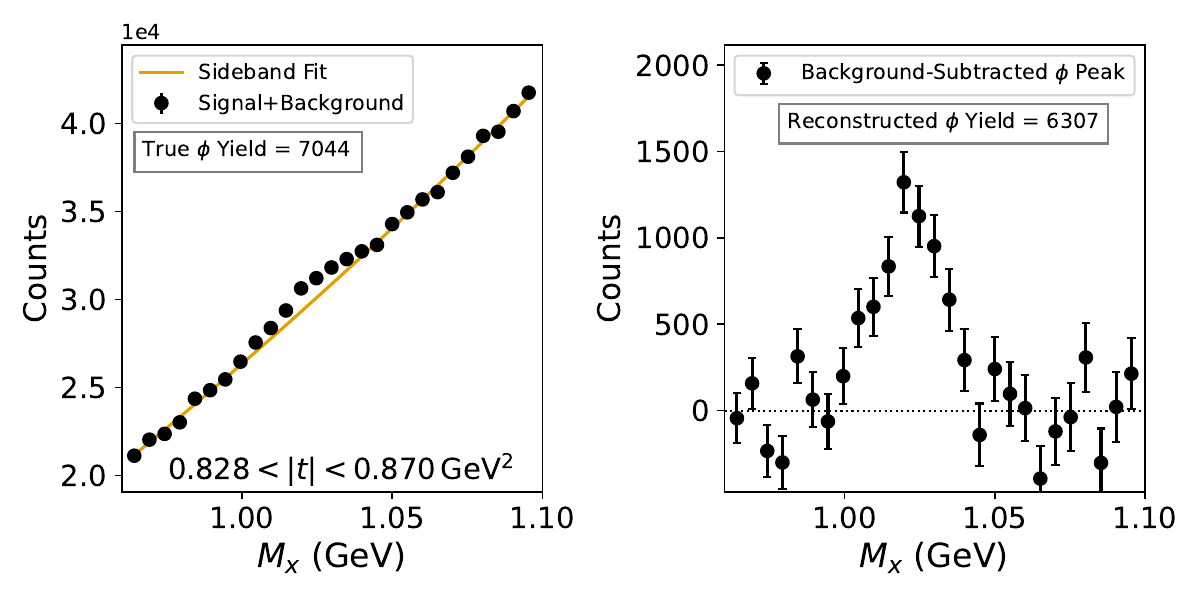}
    \label{fig:sig1}
  \end{subfigure}
  \hfill
  \begin{subfigure}[b]{0.49\textwidth}
    \includegraphics[width=\linewidth]{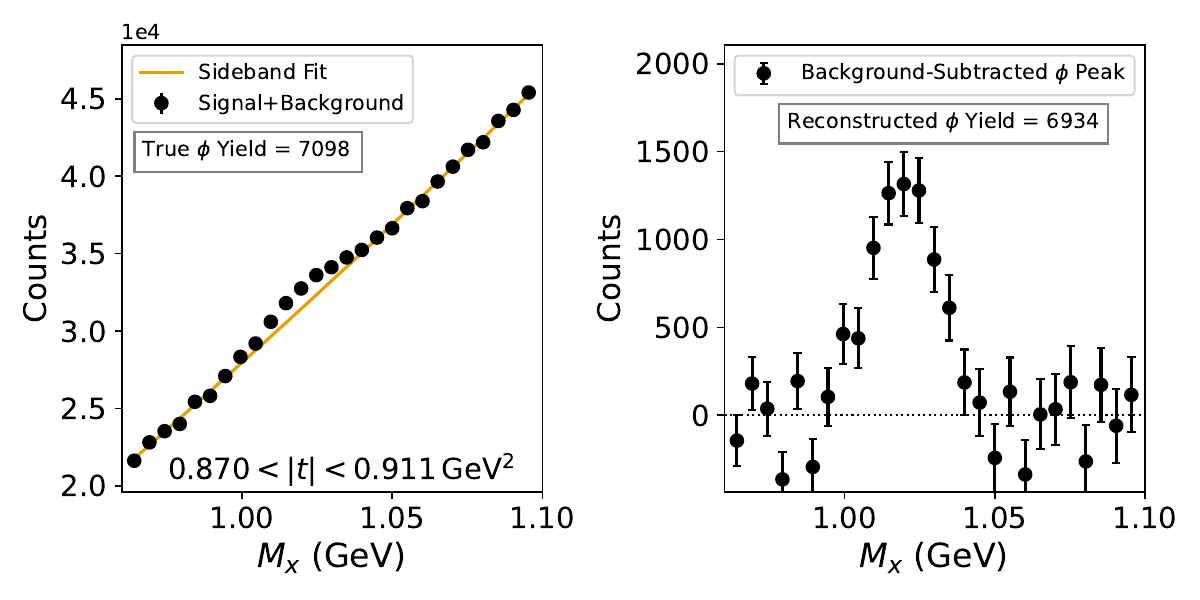}       
    \label{fig:sig2}
    \vspace{-0.25cm} 
\hfill  
  \end{subfigure}
  \begin{subfigure}[b]{0.49\textwidth}
    \includegraphics[width=\linewidth]{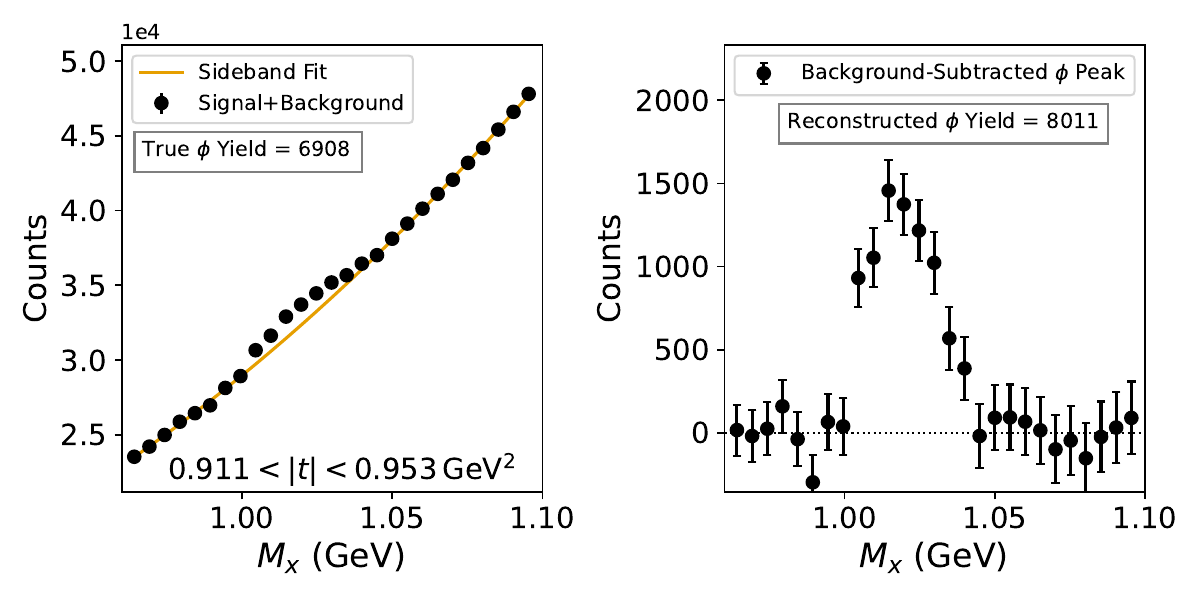}
    \label{fig:sig3}
  \end{subfigure}
  \hfill
  \begin{subfigure}[b]{0.49\textwidth}
    \includegraphics[width=\linewidth]{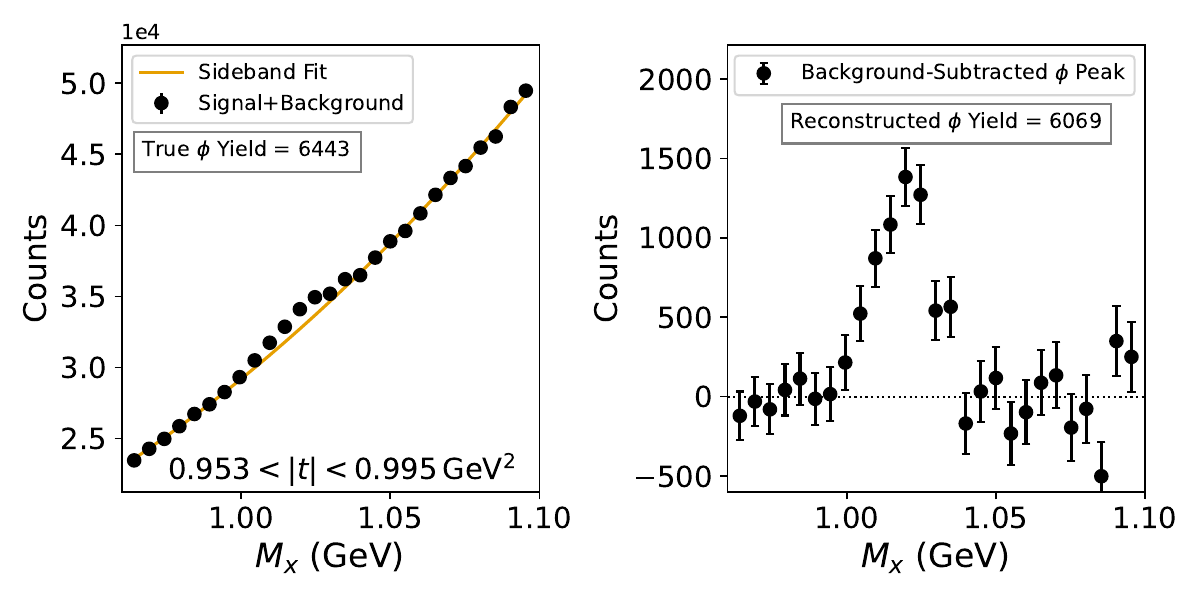}
    \label{fig:sig4}
  \end{subfigure}
  \begin{subfigure}[b]{0.49\textwidth}
    \includegraphics[width=\linewidth]{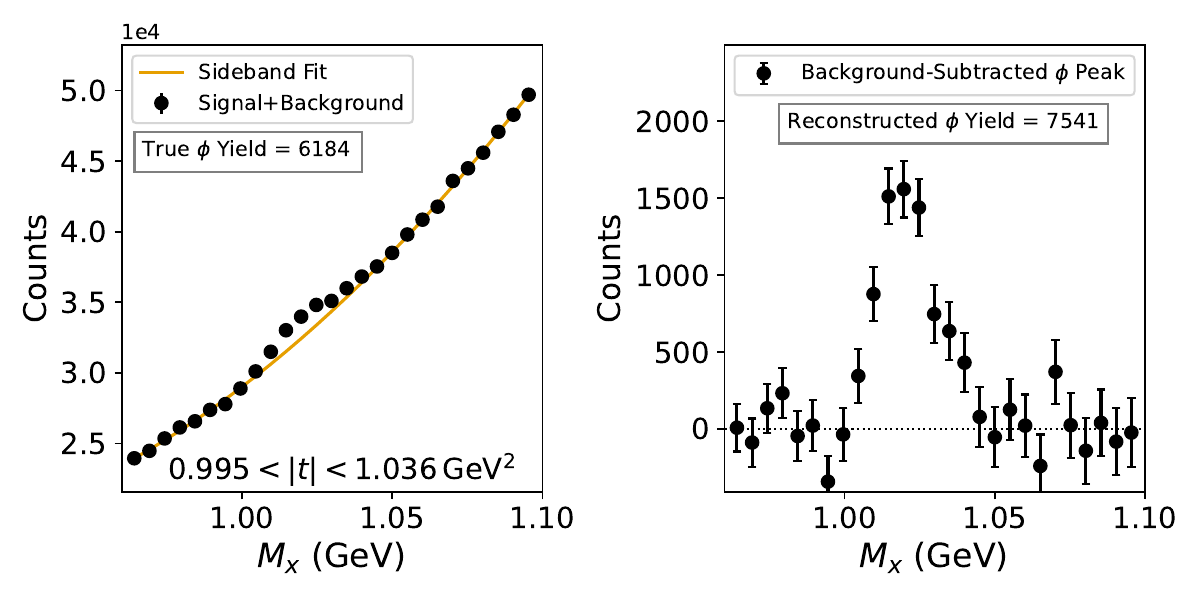}
    \label{fig:sig5}
    \end{subfigure}
     \begin{subfigure}[b]{0.49\textwidth}
    \includegraphics[width=\linewidth]{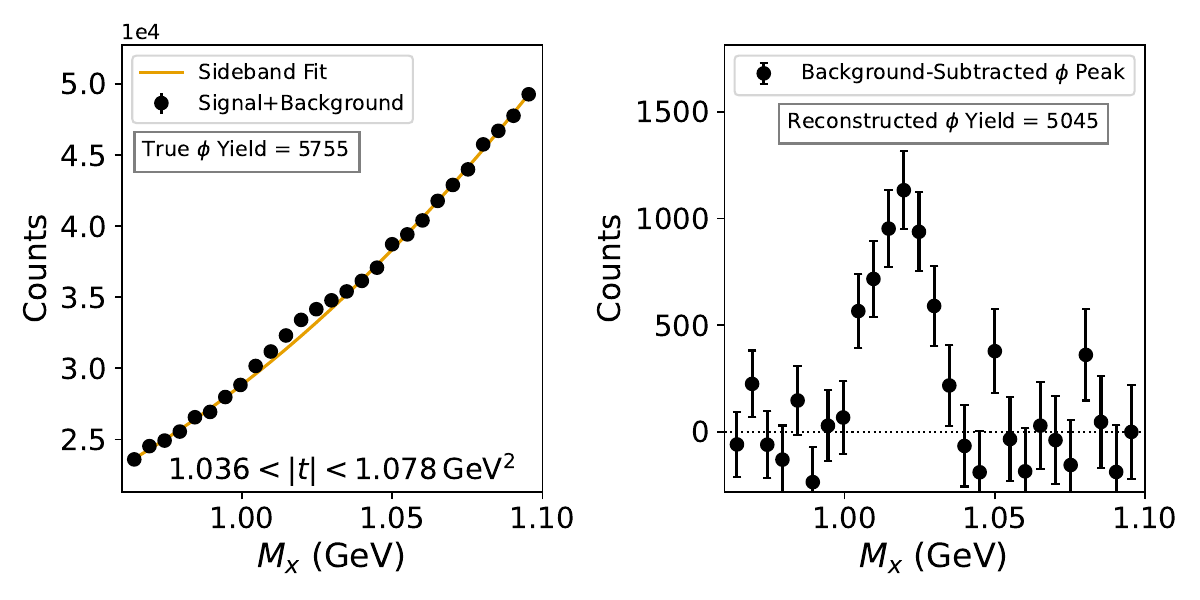}
    \label{fig:sig6}
  \end{subfigure}
  \caption{Example results of pseudo-experiments used to estimate the statistical and signal extraction uncertainty via the sideband subtraction technique. Left panels show the signal and background as it would appear in the experimental data. Right panels show the background subtracted pseudo-data distributions. These results are for $\Dszero=0.0$.}
  \label{fig:sigs}
\end{figure}

For the extraction of the $\phi$ yield in the pseudo-experiments, realistic cuts were applied on the reconstructed spectrometer variables as they would be on the real data. The set of cuts used is listed in Table~\ref{tab:SpecCuts} in the Appendix~[\ref{Sec:Appendix}].
For each \abst bin, 100,000 pseudo-experiments were performed, and the standard deviation of the yield extracted from each pseudo-experiment was taken as the combined statistical and signal extraction uncertainty. This procedure results in a point-to-point uncertainty on the $\phi$ cross section of $\approx15\%$ in each of our \abst bins. Results from a single pseudo-experiment are shown for the six \abst bins in Fig.~\ref{fig:sigs}. We tested two signal extraction techniques, the aforementioned sideband subtraction and a combined fit to the signal and background. The combined fit performed slightly better, but the two techniques provided comparable values for the uncertainties in all $|t|$ bins.

\begin{figure}[h!]
  \centering
    \includegraphics[width=0.9\linewidth]{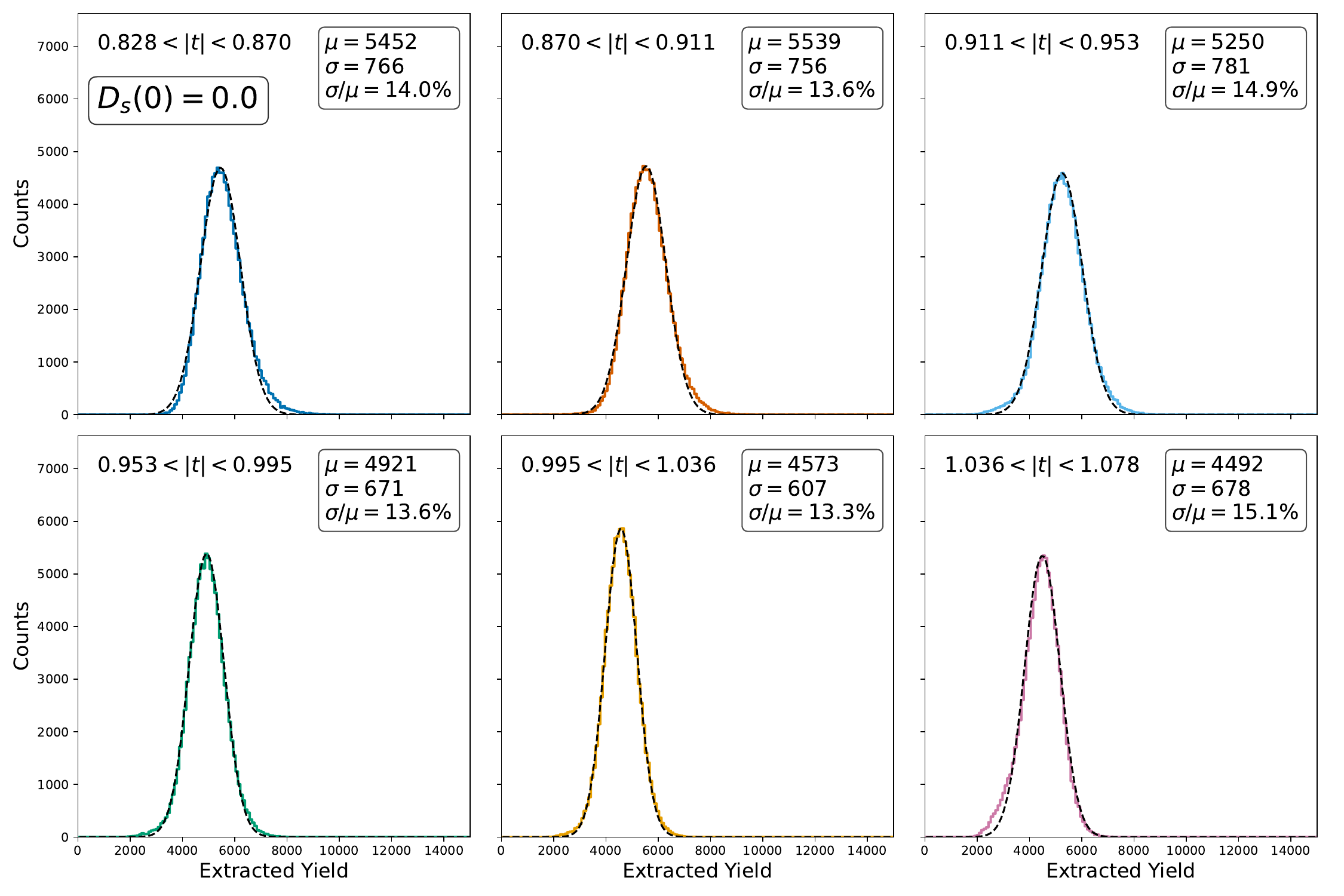}
  \caption{The measured distribution of extracted yields in each \abst bin over 100,000 pseudo-experiments. The black dashed line is a Gaussian fit to the extracted yields, the parameters of which are shown in the upper right corner of each panel. The $\sigma/\mu$ of the fit is taken as the point-to-point uncertainty from the signal extraction on that \abst bin.}
  \label{Fig:SignalExtractionUncertainty}
\end{figure}

Due to the large uncertainty on the cross section of the \etap, we additionally evaluated the impact of scaling the yield of exclusive \etap events. The signal extraction procedure was performed with 2000\% and 0\% of the \etap yield present in the \MX distribution. In all cases the extraction was able to perform similarly, with the extraction uncertainty being a few percent higher in the case of the 2000\% \etap yield. If the scaling of the \etap cross section as a function of \abst is very different than the $\phi$, there could be a bias on the resulting $\phi$ yield. We anticipate a 4\% point-to-point uncertainty on the combined modeling of the \etap and continuum background. As is the case with the signal extraction uncertainty, the background modeling uncertainty will grow if the integrated luminosity is decreased.


At the requested beam current of 75 $\mu$A, the singles rate in the HMS for our setting is expected to be high enough to cause a slight degradation in the tracking efficiency. The expected tracking efficiency in the HMS at these rates is around 97\%. The expected rates in the SHMS are relatively low, so tracking efficiencies of greater than 98\% should be achievable~\cite{Ali:2025dan}. These values are reasonably well understood, so we assign a 1\% uncertainty on the cross section arising from uncertainty in tracking efficiency.

Finally, we assign a flat 3\% point-to-point uncertainty on all points, accounting for effects such as bin centering, beam energy, proton absorption, etc., based on the experience of the VCS experiment~\cite{Li2022}.

\begin{table}[H]
\centering
\begin{tabular}{lcccccc}
\toprule
\textbf{Source} & \textbf{Bin 1} & \textbf{Bin 2} & \textbf{Bin 3} & \textbf{Bin 4} & \textbf{Bin 5} & \textbf{Bin 6} \\
\midrule
Signal Extraction        & 14.0\% & 13.6\% & 14.9\% & 13.6\% & 13.3\% & 15.1\% \\
Radiative Correction     & 4.0\%  & 4.0\%  & 4.0\%  & 4.0\%  & 4.0\%  & 4.0\%  \\
Background Modeling      & 4.0\%  & 4.0\%  & 4.0\%  & 4.0\%  & 4.0\%  & 4.0\%  \\
Tracking Efficiency      & 1.0\%  & 1.0\%  & 1.0\%  & 1.0\%  & 1.0\%  & 1.0\%  \\
Rescattering Correction  & 2.0\%  & 2.0\%  & 2.0\%  & 2.0\%  & 2.0\%  & 2.0\%  \\
Other Systematics        & 3.0\%  & 3.0\%  & 3.0\%  & 3.0\%  & 3.0\%  & 3.0\%  \\
\midrule
\textbf{Total Point-to-point} & 15.6\% & 15.2\% & 16.4\% & 15.2\% & 14.9\% & 16.6\% \\
\midrule
Acceptance Correction    & 3.0\%  & 3.0\%  & 3.0\%  & 3.0\%  & 3.0\%  & 3.0\%  \\
Value of $R$\footnotemark           & 3.8\%  & 3.8\%  & 3.8\%  & 3.8\%  & 3.8\%  & 3.8\%  \\
\midrule
\textbf{Total Normalization} & 4.8\% & 4.8\% & 4.8\% & 4.8\% & 4.8\% & 4.8\% \\
\bottomrule
\end{tabular}
\caption{Projected uncertainties on $d\sigma_L/d\abst$. The total uncertainty is calculated as the sum in quadrature of the individual uncertainties given in the table. The bin numbering scheme goes from smallest to largest $|t|$ in the range $0.828 < \abst < 1.078~\GeVsq$. Each bin has a width of 0.042 \GeVsq. }
\label{Tab:Uncertainties}
\end{table}
\footnotetext{If comparing to predictions that do not require $\sigma_L$, the normalization uncertainty on the value of $R$ can be ignored.}

\subsection{Beam Time Request}

\begin{figure}[H]
  \centering
    \includegraphics[width=0.6\linewidth]{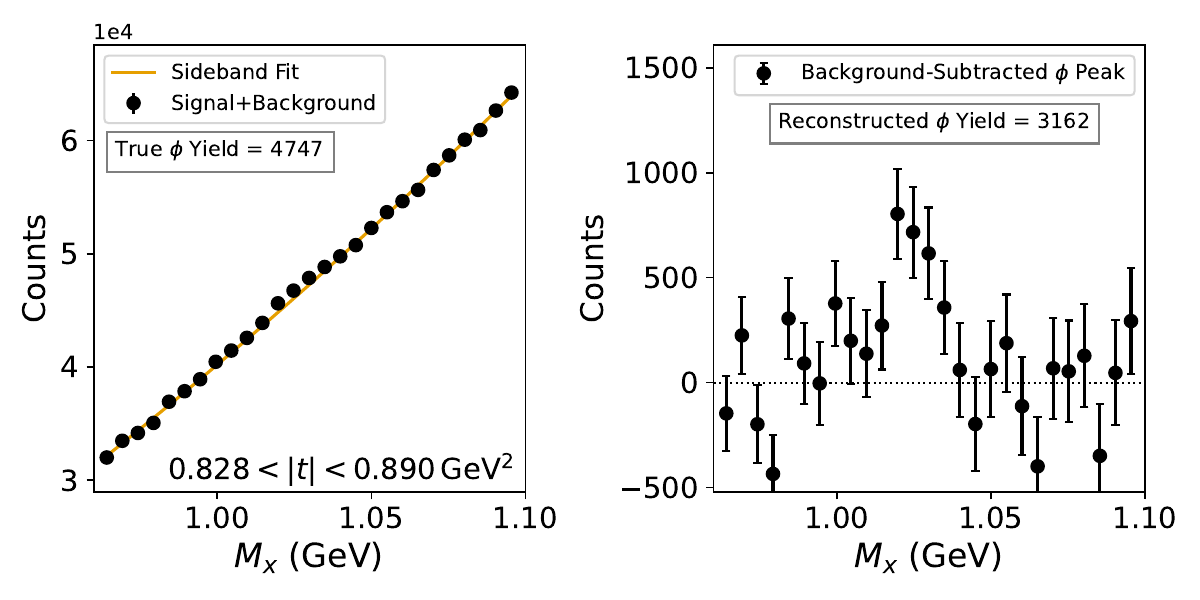}
  \caption{Example $\phi$ peak in the case of $\Dszero=-0.5$. The statistical significance of the peak is around $4\sigma$. The requirement that this peak can be resolved above background drives the requested integrated luminosity.}
  \label{Fig:PeakDs05}
\end{figure}

The desired experimental resolution on the value of $\Ds(0)$ is on the order of 0.15 to distinguish between the different hypotheses for the value of \Dszero laid out in Sec.~\ref{Sec:Theory}. The amount of beam time requested has been chosen to achieve approximately this resolution for $\Ds(0)\approx0$. The stability of the signal extraction procedure depends strongly on the integrated luminosity collected. Our requirement of 540 ab$^{-1}$ is therefore also driven by the need to reliably extract the $\phi$ peaks, particularly if the value of $\Dszero$ is of similar size to $D_{u,d}(0)$. At least four bins in $|t|$ are necessary to fit the theory to the data and extract $D_s(0)$. An example $\phi$ peak in the case $\Dszero=-0.5$ with four bins in $|t|$ is shown in Fig.~\ref{Fig:PeakDs05}. The target integrated luminosity was chosen such that peaks of $4\sigma$ significance could be obtained in four $|t|$ bins if $\Dszero=-0.5$, assuming our predictions for the $\phi$ and background cross sections are correct.

\section{Results}
\label{Sec:Results}
\begin{figure}[h]
  \centering
\includegraphics[width=0.94\linewidth ,trim=0 0.2 0 1.17cm,  
                   clip]{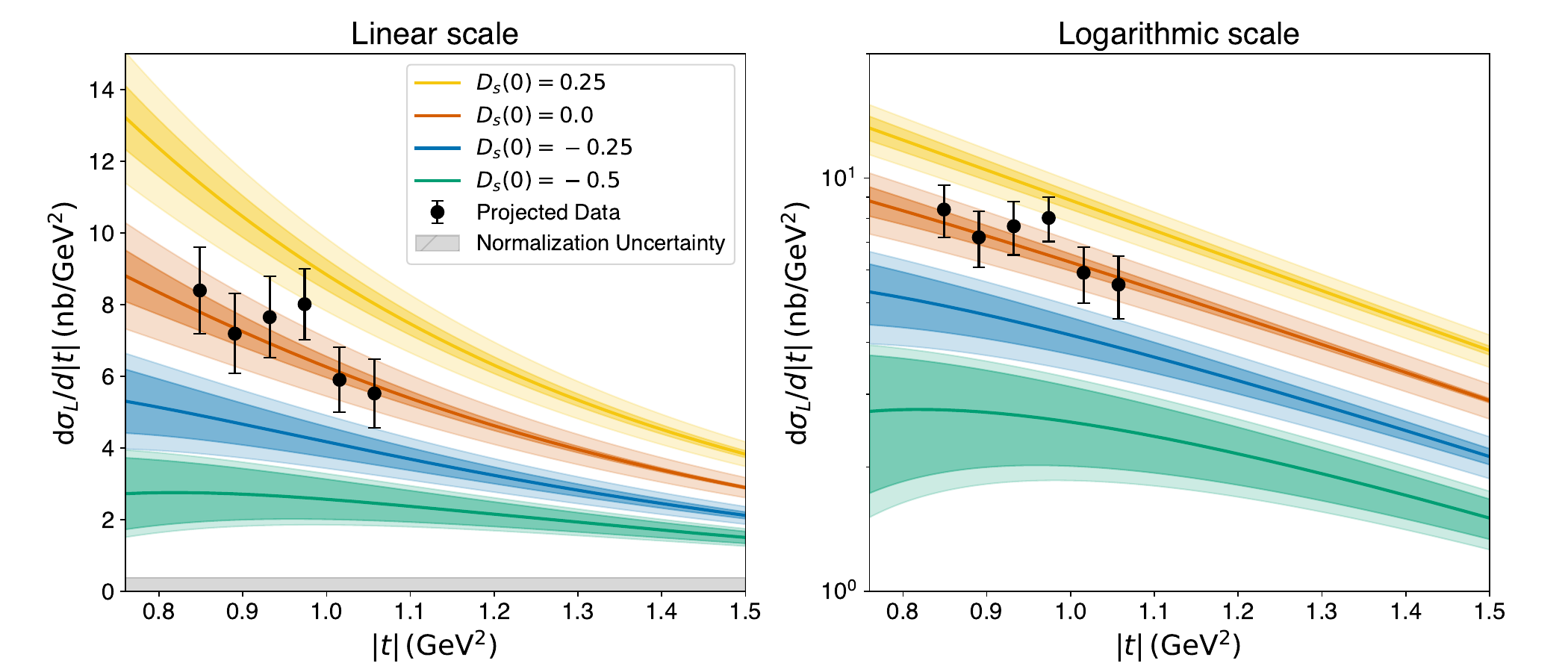}
  \caption{Projected results on $d\sigma_L/d\abst$ in linear-scale (left panel) and log-scale (right panel) for 540 ab$^{-1}$ (32 days at 75~$\mu A$). The projected data points are scattered according to their uncertainty to mimic a real measurement. There is an 4.8\% normalization uncertainty on the data.}
  \label{Fig:Projections}
\end{figure}

\begin{figure}[H]
  \centering
  \includegraphics[width=0.95\linewidth,
                   trim=0 0.2 0 0.2cm,  
                   clip
                  ]{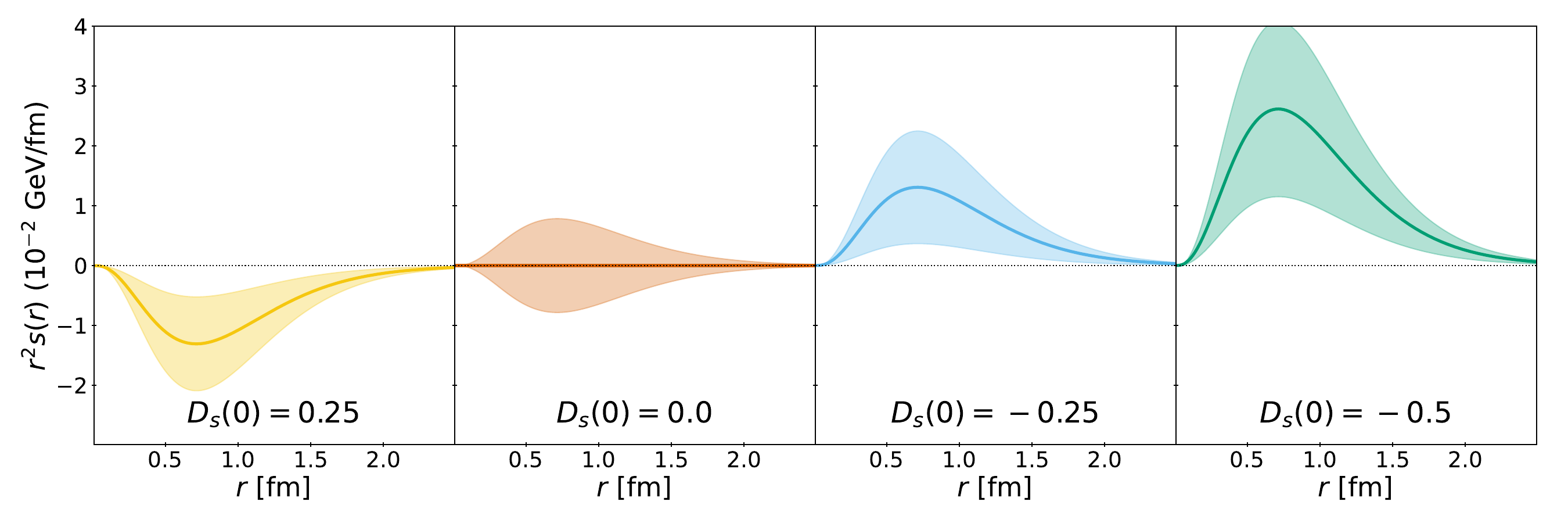}
  \caption{Example shear force distributions for four scenarios of \Dszero\ assuming the uncertainty on $\Dszero$ provided in Table~\ref{Tab:DsRes}.}
  \label{Fig:Shear}
\end{figure}

Our projected results on $d\sigma_L/d\abst$ are shown in Fig.~\ref{Fig:Projections}. To estimate the resolution on \Dszero from the projected measurement of $d\sigma_L/d\abst$, we once again employ a replica method. In this case, the data points are jittered according to their uncertainties, accounting for the degraded statistical precision at more negative values of \Dszero. The signal extraction uncertainties for the experimental data points used in this fit are shown in Appendix~\ref{Sec:SigExt}. Then the functional form of the theoretical prediction is fit to the data points, incorporating the normalization uncertainty from $R$, the uncertainty from $D_g$, and the theoretical uncertainty from the renormalization scale variation. The results are shown in Table~\ref{Tab:DsRes}. Using the expected resolutions on \Dszero provided in Table~\ref{Tab:DsRes}, in Fig.~\ref{Fig:Shear} we plot the strange quark shear force distribution for four different scenarios of \Dszero. We assume $D_s(t)$ takes the theoretically-motivated tripole form~\cite{Tong:2021ctu,Tanaka:2018wea,Lepage:1980fj} of $D_s(t)=D_s(0)(1-\frac{t}{m_D^2})^{-3}$ with the value of $m_D=0.83$ GeV determined by the most recent lattice results of Ref.~\cite{Hackett:2023rif}.
{
\setlength{\tabcolsep}{8pt}
\renewcommand{\arraystretch}{1.7}
\begin{table}[H]
 \centering
  \begin{tabular}{|c |c |c |c |c |}
    \hline
   \Dszero Value    &  0.25  &  0.0   &  -0.25 &  -0.5  \\
    \hline
    $\sigma_{\Dszero}$ & 0.15   & 0.15   & 0.18   & 0.28   \\
    \hline
  \end{tabular}
  \caption{Extracted resolutions on \Dszero\ for various values of \Dszero. At $\Dszero = 0.25$ and 0.0 the experimental uncertainty is smaller and the resolution on \Dszero is dominated by the theoretical uncertainty.}
  \label{Tab:DsRes}
\end{table}
}

\section{Other Measurement Opportunities}
\label{sec:otheropportunities}
The $\Heep X$ reaction has the experimental benefit of being ``blind" to the decay channels of the produced final state particle $X$. Thus, any particle exclusively produced in this reaction can be measured in the missing mass spectrum.  Our proposed dataset will provide a large volume of ``general-purpose" $\Heep X$ data at relatively low values of \abst. This reaction can be particularly useful for studying the production of particles with more experimentally challenging decay channels. While our primary physics focus is the $\phi$, the large momentum acceptance of the SHMS means that our proposed setting has acceptance for missing masses from roughly 0 to 1.4 GeV, as shown in Fig.~\ref{Fig:Mx}. We briefly describe in this section some of the extra measurements that could be made using this data with no additional runtime or changes to the experimental setup.

\subsection[Measurement of omega]{Measurement of $\omega$}

Exclusive leptoproduction of $\omega$ mesons has been studied in a variety of experiments~\cite{Joos:1977tz,CLAS:2005nkx,ZEUS:2000swq,HERMES:2014xob,HERMES:2015zqh}, due in part to the connection between DVMP and the GPDs. At present, most attempts to fit and extract GPDs utilize only DVCS data, which poses a difficult inverse problem~\cite{Bertone:2021yyz,Moffat:2023svr}. DVMP provides a complementary process which can help constrain GPDs. In addition to measurements of absolute cross sections, ratios of exclusive vector mesons can be predicted by GPD models~\cite{Diehl:2003ny} and quark charge counting rules~\cite{Ivanov:2004ax}. For this reason, vector meson cross section ratios have been measured at a variety of energies~\cite{H1:2020lzc,H1:2000hps,ZEUS:2005bhf,Laget:2019tou}. Studying the ratios of the $\omega$ and $\phi$ production cross sections could aid the primary measurement by helping to understand both the applicability of GPD models in the near-threshold kinematic region and the interplay of the quark and gluon contributions to the DVMP cross section. 

The authors of~\cite{Wang:2021dis} used near-threshold photoproduction data of $\omega$ and $\phi$ to extract the proton mass radius and reached similar results to the mass radius determined using $J/\psi$ data, in spite of the fact that no hard scale is present and DVMP factorization should not hold. Our data on both $\phi$ and $\omega$, with the hard scale set mostly by $Q^2$, can serve as a cross check of this result. 


The broad mass distribution of the $\rho$ means it has a substantial overlap with continuum processes, meaning an absolute cross section measurement of $\rho$ is unlikely with our dataset. Since the $\omega$ is much narrower, it can be reasonably well separated from the $\rho$ and other processes. This separation has been demonstrated with a similar spectrometer setup in the $u$-channel $\omega$ measurement of Refs.~\cite{JeffersonLabFp:2019gpp,Li:2017xcf}. The study of Ref.~\cite{JeffersonLabFp:2019gpp} found that in the $u$-channel reaction the peak near the $\omega$ mass was predominantly $\omega$, with a smaller continuum contribution of $\rho$ and two-pion production. In lieu of a dedicated background sample of $\rho$ and two-pion production, we simply perform a sideband subtraction on the peak of the $\MX$ distribution located at the $\omega$ mass and assume the extracted yield is purely $\omega$. We assign an 8\% systematic uncertainty related to the subtraction of the $\rho$ and two-pion backgrounds. The systematic uncertainties from Table~\ref{Tab:Uncertainties} are also applied, bringing the total systematic uncertainty on all points to 10.5\% as a rough estimate.

\begin{figure}[h]
  \centering
    \includegraphics[width=0.9\linewidth]{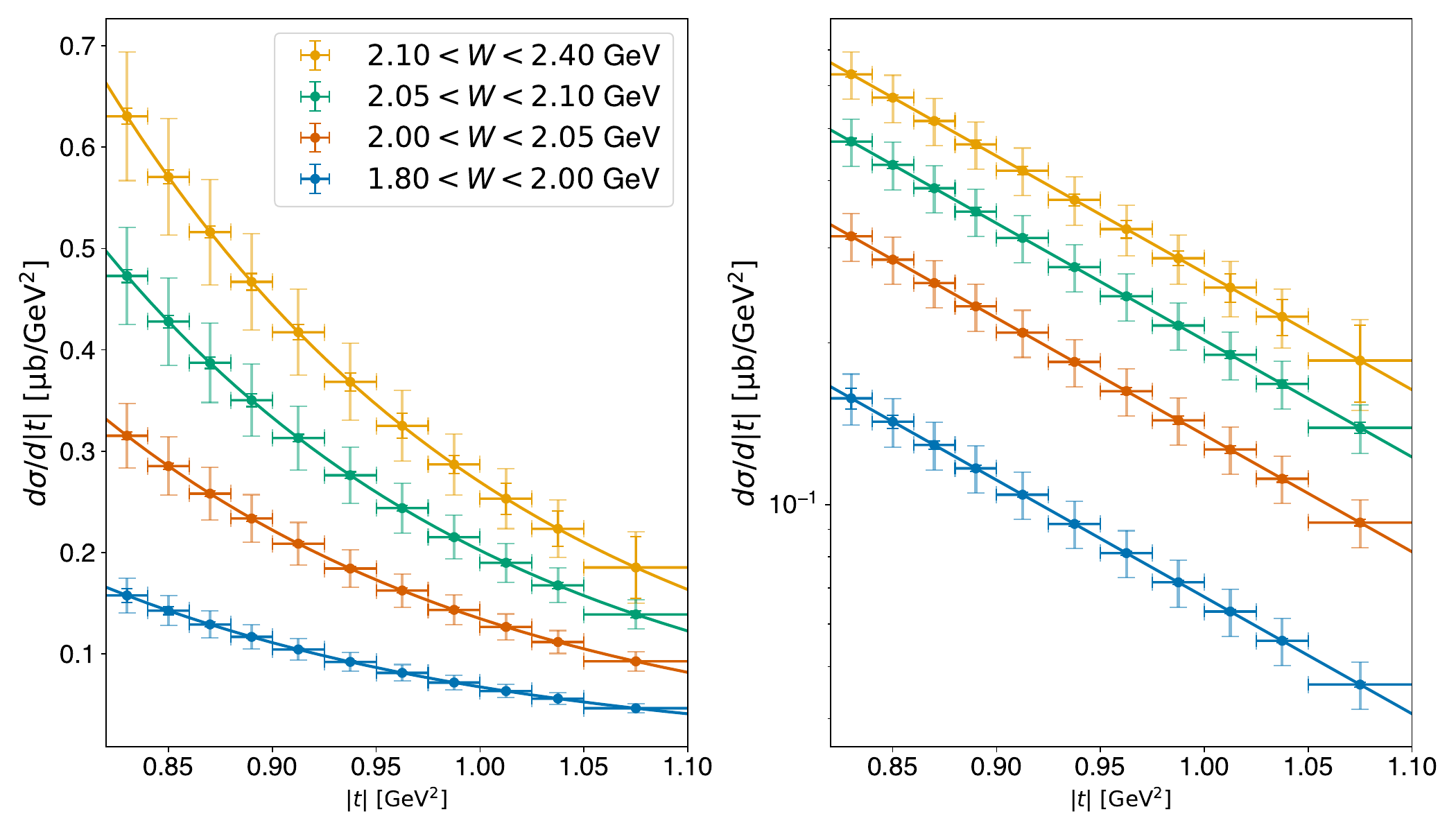}
  \caption{Projected measurement of the $\omega$ electroproduction cross section as a function of \abst and $W$. The inner error bar is statistical and the outer is the statistical and systematic added in quadrature. The lines correspond to a simple exponential $|t|$-distribution. }
  \label{Fig:omegaXsec}
\end{figure}

Based on the results shown in Fig.~\ref{Fig:omegaXsec}, it is clear that this dataset provides a promising avenue to study $t-$channel $\omega$ electroproduction. In the scenario where the $\phi$ cross section is smaller than predicted and cannot be resolved, a differential measurement of $\omega$ production remains a concrete deliverable of our proposed experiment.

\subsection[Measurement of etap]{Measurement of $\eta'$ and $\eta$}

\begin{figure}[H]
  \centering
    \includegraphics[width=0.7\linewidth]{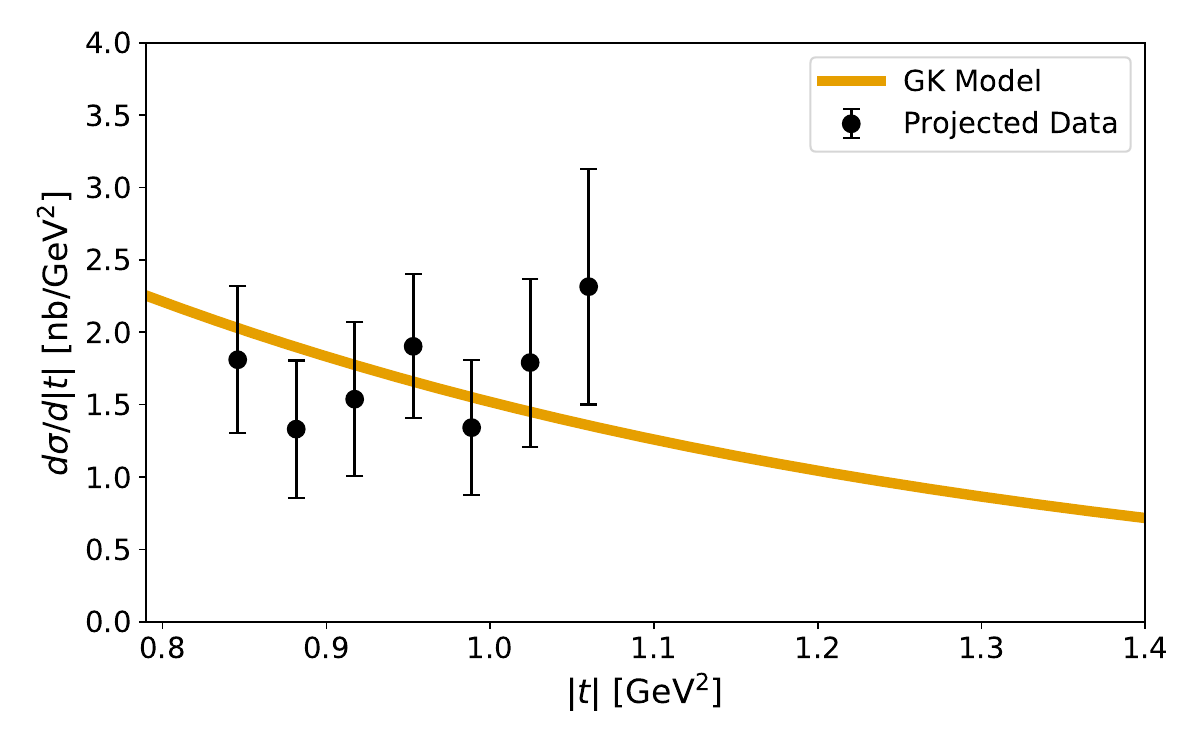}
  \caption{Projected measurement of the \etap electroproduction cross section as a function of \abst. The uncertainties are dominated by the signal extraction uncertainty. The datapoints are jittered randomly according to their uncertainties. The theory curve is from Refs.~\cite{Goloskokov:2011qa,Duplancic:2023xrt} and incorporates the full twist-3 contribution.}
  \label{Fig:EtapXsec}
\end{figure}

With the data collected by this proposed experiment, the first ever measurement of \etap~electroproduction could be performed\footnote{There was a proposal made to PAC16 (PR-99-109) to perform a measurement of \etap~electroproduction with CLAS; however, the results were never published.}. The \etap~is unique due to its position as a near-flavor-SU(3)-singlet amongst the pseudoscalar mesons and to its unexpectedly large mass, which is generated by the QCD chiral anomaly~\cite{tHooft:1976rip}. There exist GPD predictions for the \etap cross section~\cite{Goloskokov:2011rd}, which we compare to our projected data in Fig.~\ref{Fig:EtapXsec}. Our projections employ the same replica technique as was used for the $\phi$ in Sec.~\ref{Sec:Exp} to estimate the systematic uncertainties\footnote{Since we did not have a dedicated signal sample of generated \etap events, our extraction procedure could not maximally utilize a realistic prior for the shape of the reconstructed \etap distribution. For this reason, the uncertainties are larger than for the $\phi$, although they will likely shrink once a dedicated sample of \etap is generated and studied.}.

\begin{figure}[H]
  \centering
    \includegraphics[width=0.7\linewidth]{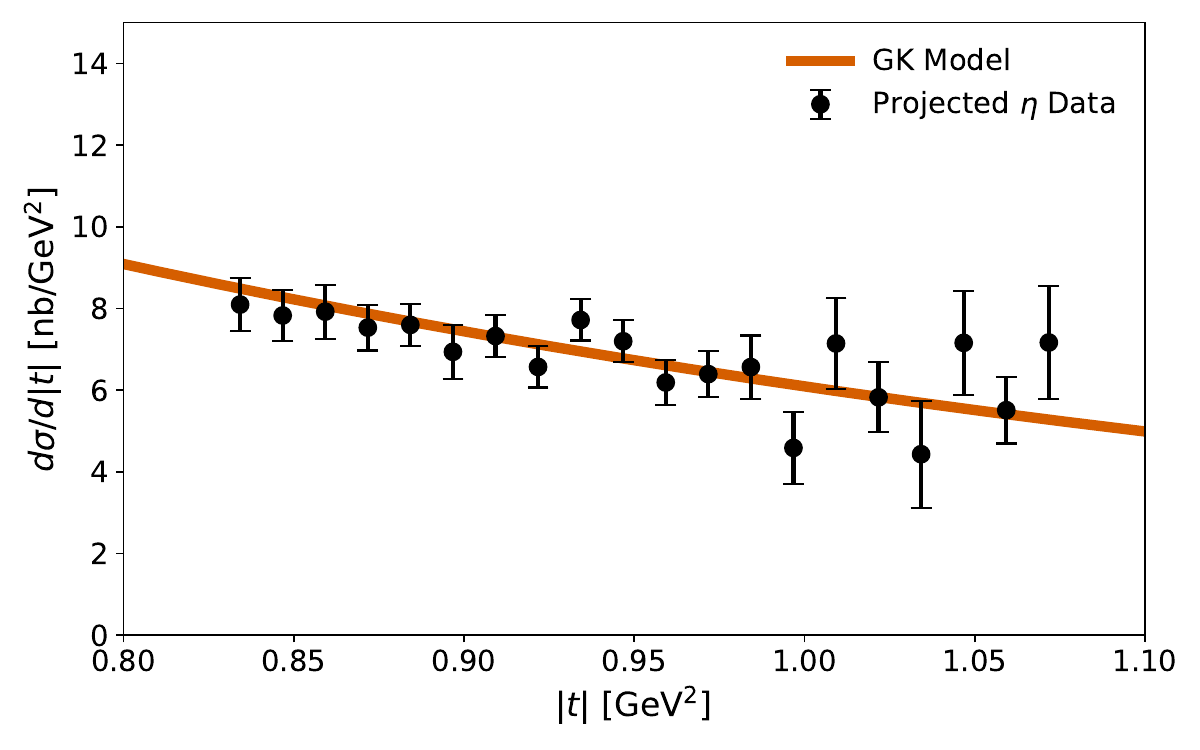}
  \caption{Projected measurement of the $\eta$ electroproduction cross section as a function of \abst. The datapoints are jittered randomly according to their uncertainties.}
  \label{Fig:EtaXSec}
\end{figure}

In addition to the first measurement of \etap electroproduction, we can perform a highly differential measurement of exclusive $\eta$ electroproduction, shown in Fig.~\ref{Fig:EtaXSec}. We treat the uncertainties in the same way as for the $\etap$ and $\phi$. Since the $\eta$ cross section is larger and the background is flatter than under the $\etap$ and $\phi$, the uncertainty associated with the signal extraction is significantly smaller in most bins. Similarly to the $\omega$, the $\eta$ cross section is a concrete deliverable of our experiment in the event that the $\phi$ cross section is too small to measure.

Performing these measurements of \etap and $\eta$ electroproduction would provide valuable information for GPD modeling and extraction. The CLAS collaboration studied electroproduction of the pseudoscalar mesons $\eta$~\cite{CLAS:2017jjr} and $\pi^0$~\cite{CLAS:2014jpc} at $W>2$ GeV to access the chiral-odd transversity GPDs. There is also a proposal (E12-06-108) to perform similar measurements with CLAS12. The different quark flavors of the $\pi^0$ and $\eta$ enabled study of the flavor dependence of these GPDs. Historically it has been assumed  that the strangeness of the $\eta$ plays no role. Future measurements of hard exclusive production of \etap and $\eta$, which differ in their strange quark content, can test this assumption. The authors of Ref.~\cite{Goloskokov:2011rd} computed the total \etap and $\eta$ cross sections, as well as the separated $\sigma_L$ and $\sigma_T$ and beam-spin asymmetries, using a GPD model. Interestingly, the total $\eta$ cross section is predicted to be dominated by $\sigma_T$ while the \etap cross section is predicted to be predominantly $\sigma_L$. We show our projected data uncertainties compared to this model in Figs.~\ref{Fig:EtapXsec} and \ref{Fig:EtaXSec}\footnote{We received numerical predictions from the authors of Ref~\cite{Goloskokov:2011rd} only for the $\etap$; to generate the predictions for the $\eta$ shown in Fig.~\ref{Fig:EtaXSec} we simply scaled the $\etap$ predictions up by a factor corresponding roughly to the predicted ratio of the $\eta$ and $\etap$ cross sections.}. Our data can provide a reasonable consistency check of GPD models at these kinematics. A measurement of the size and \abst-distribution of the \etap~cross section could further inform future dedicated measurements targeting the quark-flavor-dependent transversity GPDs. 

In addition to GPDs, the non-perturbative gluon content of the $\eta$ and \etap has long been an active field of study, motivated by the possibility of mixing with glueball states~\cite{Robson:1977pm,Shore:1991np,Ball:1995zv,Bass:2008fr,Escribano:2007cd,Ochs:2013gi}. These states are also accessible to lattice computations~\cite{Gregory:2011sg,Fukaya:2015ara,Simeth:2022fuq}. The authors of Ref.~\cite{Kroll:2002nt} proposed electroproduction of $\eta$ and \etap mesons as a probe of their two-gluon Fock components. Although the kinematics of our experiment are not perfectly aligned with the preferred kinematics of Ref.~\cite{Kroll:2002nt}, it would nevertheless be interesting to compare their predictions to our data.

\subsection{Beam Spin Asymmetries}
Throughout the above discussion of proposed measurements, the polarization of the beam was not mentioned. Having some degree of longitudinal polarization of the electron beam for our experiment would enable us to access an interesting set of beam-spin asymmetries (BSAs). We want to stress that polarization is not a requirement for scheduling our experiment, but having it would extend the physics reach of our dataset.

The beam-spin asymmetries for DVMP of $\phi$, \etap, $\eta$, and $\omega$ are all accessible with this dataset and provide useful information for understanding GPDs and testing GPD models. The magnitude of the BSA is \begin{eqnarray}\label{eq:BSA}
	\text{BSA} = \frac{\sqrt{2 \epsilon (1 - \epsilon)} \frac{\sigma_{LT^{\prime}}}
		{\sigma_{0}}\sin\phi_h}
	{1 + \sqrt{2 \epsilon (1 + \epsilon)}\frac{\sigma_{LT} }{\sigma_{0}} \cos\phi_h
		+ \epsilon \frac{\sigma_{TT}}{\sigma_{0}} \cos2\phi_h},
\end{eqnarray}
where $\sigma_{0}=\sigma_T+\epsilon\sigma_L$. $\sigma_{LT}$, $\sigma_{TT}$, and the polarized term $\sigma_{LT^\prime}$ describe the interference between the amplitudes of $\sigma_L$ and $\sigma_T$. In fact, this BSA term can, in conjunction with theoretical inputs or other experiments, help constrain the magnitude of terms that could affect the extraction of $\sigma_L$. Unfortunately, mapping a large region of $\varphi_h$ is infeasible with our dataset, but we can provide results at a fixed value of $\varphi_h$ that can be used and compared to theoretical predictions and measurements from larger acceptance experiments.

\section{Summary}
We propose to measure $d\sigma/d|t|$ for $\phi$ electroproduction at $W\approx2.25$ GeV and $\Qsq\approx3.4$ GeV$^2$ via the $\Heep \phi$ reaction in Hall C. Our measurement will provide the first constraints on the strangeness $D$-term of the proton, in particular on the magnitude of \Dszero, which will have a high impact on our understanding of the mechanical structure of the proton.
Historically, strangeness has been ignored in most studies of the proton's GFFs. Understanding the effects of strangeness will help eliminate assumptions that have been made in the extraction of the proton’s mechanical quantities.
If our measurements reveal that the magnitude of \Dszero is large, we will establish strangeness as an important contributor to the proton's mechanical structure. 
Conversely, values of \Dszero near zero would confirm $\phi$ electroproduction as a promising process to access the \textit{gluon} $D$-term, supporting a gluonic interpretation for other experimental efforts to measure exclusive $\phi$ production at Jefferson Lab energies.
In either case, the results from our proposed experiment will add important new constraints to future global fitting efforts, and will impact all studies of the proton's total $D$-term. 
Furthermore, our experiment will provide a large, multi-purpose dataset of $\Heep X$ data that touches on topics reaching from proton GPDs to the gluonic structure of hadrons, as outlined in Sec.~\ref{sec:otheropportunities}, including the first measurement of \etap electroproduction.
These results on exclusive $\phi$, $\omega$, $\eta$, and $\eta'$ electroproduction will provide a testing ground for hadronic and partonic descriptions for electron scattering off protons. 

To achieve these results, we request 32 days of unpolarized electron beam at 10.6 GeV with a current of 75 $\mu A$ on the standard Hall C 10 cm LH$_2$ target, along with three days for optics calibration and commissioning, bringing the total request to 35 days. 

\clearpage
\section{Comparison to Other Experiments and Proposals}

The CLAS experiment performed two measurements of exclusive $\phi$ electroproduction~\cite{CLAS:2001zwd,CLAS:2008cms} during the 6 GeV era (E-93-022, E-99-105), with beam energies of 4.2 GeV and 5.754 GeV. In both cases, the cross sections were measured as a function of $t'=|t-t_{min.}|$ and \Qsq. The range in \Qsq of the measurement at 4.2 GeV was 0.7 to 2.2 \GeVsq. The measurement at 5.7 GeV extended the kinematic range to $1.4<\Qsq<3.8$~\GeVsq. The 5.7 GeV measurement additionally presents the cross section $d\sigma/d\abst$, which is the necessary quantity for extraction of \Dszero. However, as previously mentioned, the authors of the theory prediction were unable to use these results to extract \Dszero, as the measured differential cross section $d\sigma/d\abst$ was integrated over the whole kinematic phase space of the measurement, i.e., $1.4\leq\Qsq\leq3.8~\GeVsq$ and $2.10 \leq W \leq 2.90$ GeV. The CLAS $d\sigma/d\abst$ is thus dominated by events at lower \Qsq and higher $W$ than the data that will be collected as a part of this experiment.

The existing 12 GeV proposal most similar to ours was submitted to PAC 39 under the title ``Exclusive Phi Meson Electroproduction with CLAS12" (PR12-12-007)~\cite{CLAS12Phi}. One of the primary goals of this proposal is to measure the \abst-slope at low \abst and large $W$ for extraction of the high-$x$ gluon GPD. The goal of our study is complementary but different. For the extraction of \Dszero as discussed in Sec.~\ref{Sec:Theory}, the \abst-distribution at low \abst must be measured near-threshold at large $\xi$ in as small of a window in \Qsq and \W~as possible. This requirement is uniquely befitting a high luminosity spectrometer setup, where high event statistics can be collected in narrow regions of phase space. With the integrated luminosity of CLAS12 Run Group A, a measurement that satisfies all the kinematic constraints of the theory laid out in Sec.~\ref{Sec:Theory} is not possible. However, an $L/T$ separation of the exclusive $\phi$ cross section is foreseen via measurement of the $\phi$ spin density matrix elements in the CLAS12 proposal, and that data will be useful for interpreting our result. Similarly, our results on the effect of strangeness on $\phi$ production will aid in the physics interpretation of the CLAS12 results. Given that our measurement and the CLAS12 measurement will employ independent detectors and analysis techniques, the two measurements will cross check and enhance the validity of each other's results.

The CLAS12 ALERT run group has proposed measurements of exclusive $\phi$ electroproduction on helium-4 and deuterium targets (E12-17-012C)~\cite{Armstrong:2017wfw}. The comparison of the $\phi$ electroproduction cross section between proton and nuclear targets is clearly interesting for a multitude of reasons, not the least of which is the fact that helium-4, as the simplest spin 0 system, has non-zero contributions from only the $A$ and $D$ GFFs. While a comparison between proton and nuclear targets directly within CLAS12 is certainly possible, the data we propose to collect will provide an additional proton target reference dataset for the ALERT $\phi$ program.

The E12-23-004 experiment entitled ``A Search for a Nonzero Strange Form Factor of the Proton at 2.5 $\mathrm{(GeV/c)^2}$"~\cite{NonzeroStrange} seeks to measure the strange form factor of the proton via parity-violating elastic scattering. A better knowledge of the contribution of strangeness to the electromagnetic structure of the proton would enhance the interpretation of our proposed results, and, taken together, this study and ours will together fit nicely into the larger JLab program of studying strangeness in the proton.

An LOI for a measurement of $\phi$ electroproduction in Hall A utilizing the HRS and SBS was submitted to PAC 35 (LOI-10-002). In this case, the physics goals were similar to those of the two CLAS12 proposals mentioned above; however, a full proposal was not submitted.

\section{Acknowledgements}
We would like to sincerely thank J. Schoenleber, K. Passek-K., P. Kroll, and S. Goloskokov for providing theory calculations and for valuable discussion and comments. We also thank L. DeWitt for editing this manuscript.

\clearpage
\appendix
\section{Appendix}
\label{Sec:Appendix}
\subsection{Cross Section Parameterization}
We utilize in the above proposal the $\phi$ electroproduction parameterization from Ref.~\cite{CLAS12Phi}. This parameterization was fit to the existing data on $\sigma_T, \sigma_L,$ and $R = \frac{\sigma_L}{\sigma_T}$ from Cornell, H1, ZEUS, HERMES, NMC, and CLAS. The functional form of the transverse cross section is given as:
\begin{equation}
\sigma_T(W,\Qsq) = \frac{c_T(W)}{(1+\Qsq/m_\phi^2)^{\nu_T}},
\end{equation}
where $\nu_T$ is 3.0 and
\begin{equation}
c_T(W) = \alpha_1(1-\frac{W^2_{\mathrm{Th.}}}{W^2})^{\alpha_2}(\frac{W}{\mathrm{GeV}})^{\alpha_3}
\end{equation}
in units of nanobarns. $W_{\mathrm{Th.}} = 1.96$ GeV, $\alpha_1 = 400, \alpha_2 = 1.0,$ and $\alpha_3 = 0.32$. The longitudinal cross section is given as
\begin{equation}
\sigma_L(W,\Qsq) = R(W,\Qsq)\sigma_T(W,\Qsq),
\end{equation}
where 
\begin{equation}
R(W,\Qsq) = \frac{0.4\cdot\Qsq}{m^2_{\phi}}.
\end{equation}
The authors additionally provide two models for the \abst dependence of the cross section, an exponential and a dipole. We provide only the dipole model that we used in our proposal here for compactness. The cross section differential in \abst is given by
   \begin{equation} 
    \frac{d\sigma_{L,T}}{d\abst} = \frac{\sigma_{L,T}F(\abst)}{F_{\mathrm{Int.}}},
    \end{equation}    
   where 
    \begin{equation}
   F(\abst) = \frac{m_g^8}{(m_g^2-t)^4}
    \end{equation}
    and
    \begin{equation}
   F_{\mathrm{Int.}} = \frac{m_g^8}{3(m_g^2-t_{\mathrm{min.}})^3}.
    \end{equation}  
   The mass $m_g$ is chosen to be 1.2 GeV.
   
\subsection{Signal Extraction Results at Different $D_s$}
\label{Sec:SigExt}
While the results in Fig.~\ref{Fig:Projections} are for the nominal value of $D_s(0)$, we also study the stability of the signal extraction at larger negative values, where the cross section is smaller. The length of the requested beamtime is such that we should be able to reliably perform a measurement of \Dszero down to at least a \Dszero of -0.5.

\begin{figure}[H]
  \centering
    \includegraphics[width=0.95\linewidth]{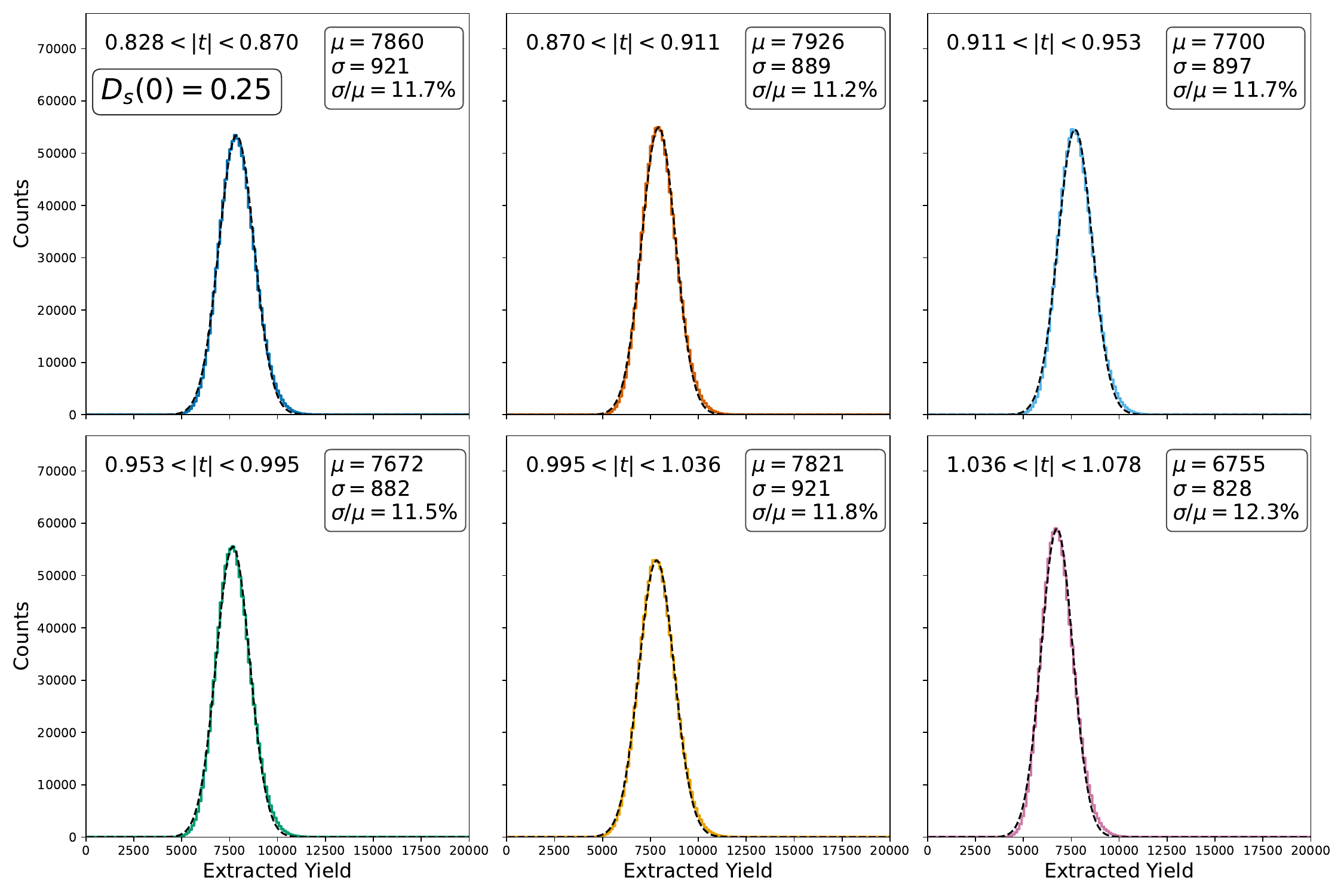}
  \caption{Extracted yield of $\phi$ events for $\Dszero = 0.25$. The cross section for $\phi$ production is increased by a factor of 1.35 compared to $\Dszero=0.0$.}
  \label{Fig:ExtractionDs025}
  \end{figure}

\begin{figure}[H]
  \centering
    \includegraphics[width=0.95\linewidth]{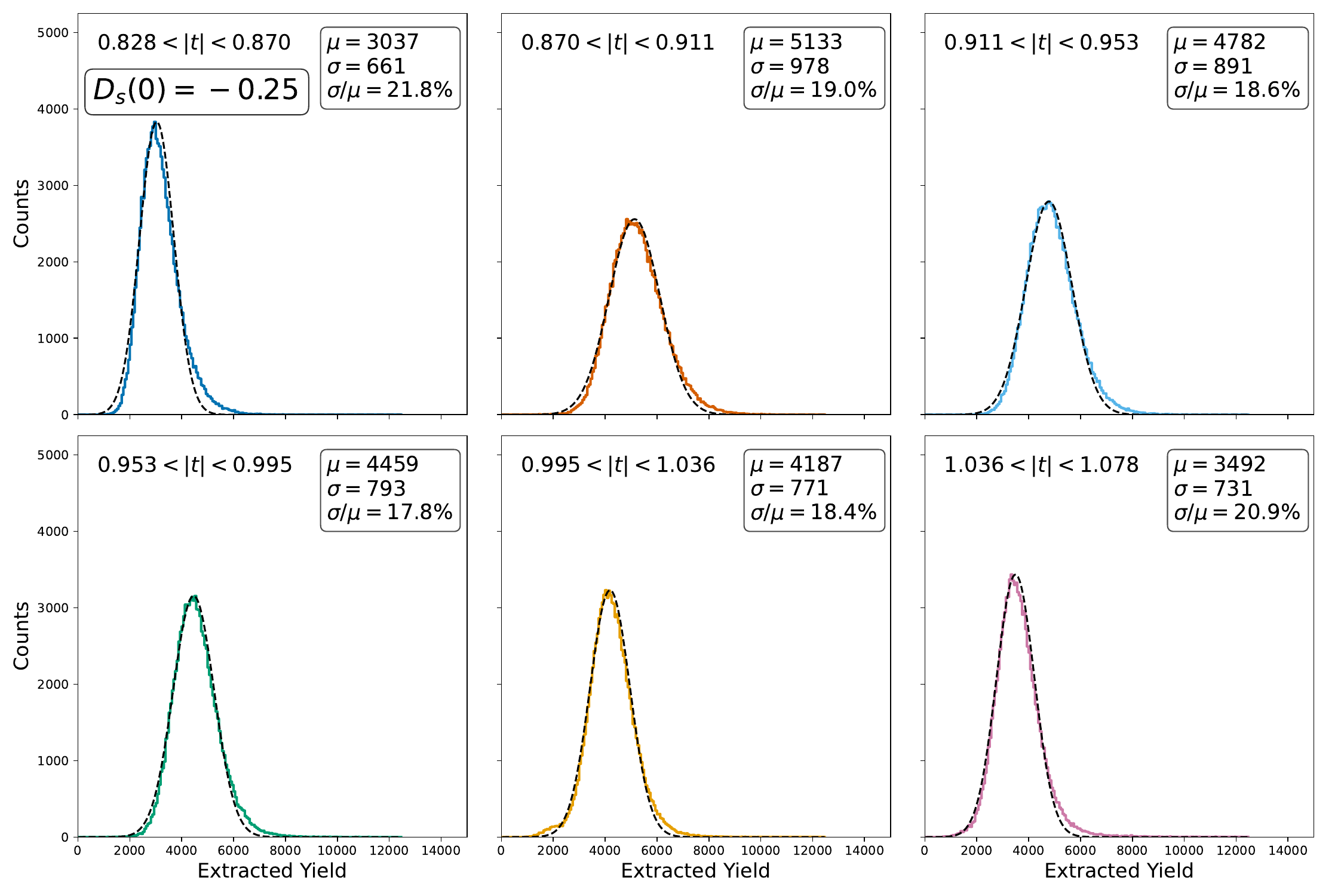}
  \caption{Extracted yield of $\phi$ events for $\Dszero = -0.25$. The cross section for $\phi$ production is reduced by a factor of 1.44 compared to $\Dszero=0.0$.}
  \label{Fig:ExtractionDs025}
  \end{figure}

  \begin{figure}[H]
  \centering
    \includegraphics[width=0.63\linewidth]{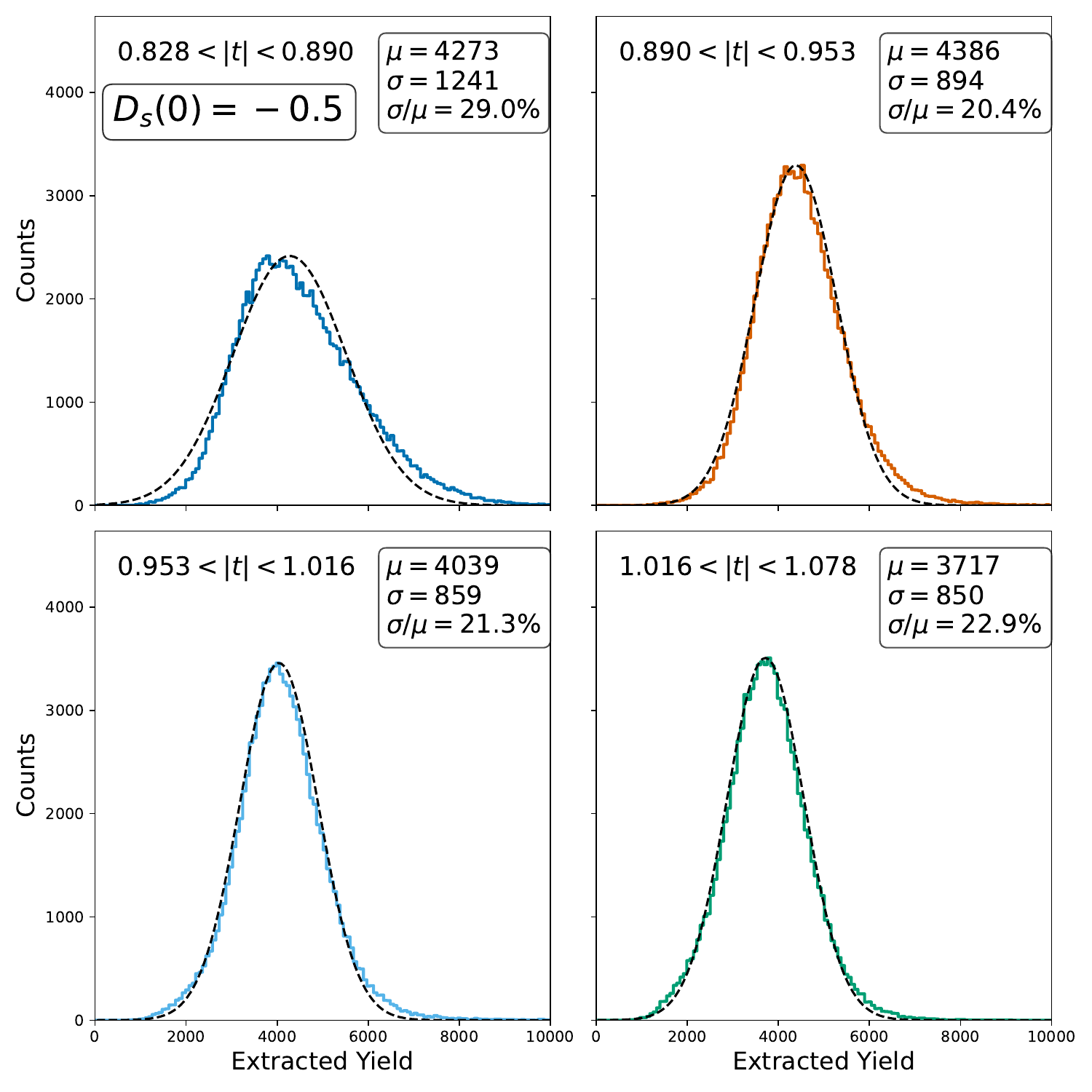}
  \caption{Extracted yield of $\phi$ events for $\Dszero = -0.5$. The cross section for $\phi$ production is reduced by a factor of 2.26 compared to $\Dszero=0.0$. Some bins exhibit a non-gaussian probability distribution. In principle, the shape of this distribution can be included in the fit for extracting \Dszero. However, for the purposes of this proposal we simply use the $\sigma$ of the gaussian fit.}
  \label{Fig:ExtractionDs05}
\end{figure}


  \subsection{Spectrometer Cuts}
The cuts applied on our simulated data are fairly standard cuts on the spectrometer acceptances in polar angle ($y'_{\text{tar.}}$), azimuthal angle ($x'_{\text{tar.}}$), and momentum ($\delta$). 
\begin{table}[h!]
  \centering
  \begin{tabular}{lccc}
    \hline
     & $\delta$ (\%) & $x'_{\text{tar.}}$ & $y'_{\text{tar.}}$ \\
    \hline
    HMS  & $-8 < \delta < 8$ & $-0.08 < x'_{\text{tar.}} < 0.10$ & $-4 < y'_{\text{tar.}} < 4$\\
    SHMS & $-10 < \delta < 22$ & $-0.06 <x'_{\text{tar.}} < 0.06 $& $-4 < y'_{\text{tar.}} < 4$ \\
    \hline
  \end{tabular}
  \caption{Kinematic variables and the applied cuts for the HMS and SHMS.}
  \label{tab:SpecCuts}
\end{table}

\clearpage
\bibliographystyle{unsrt}
\bibliography{refs}

\end{document}